\documentclass[epj]{svjour}
\usepackage{latexsym}
\usepackage{graphics}
\usepackage[latin1]{inputenc}
\usepackage{rotating}
\begin{document}
\renewcommand{\textfraction}{0.00000000001}
\renewcommand{\floatpagefraction}{1.0}
\title{Photoproduction of $\eta '$-mesons off the deuteron}
\author{ 
        I.~Jaegle\inst{1},
        T.~Mertens\inst{1},
	A.~Fix\inst{8},
	F.~Huang\inst{9},
	K.~Nakayama\inst{9},
	L.~Tiator\inst{10},
        A.V.~Anisovich\inst{2,3},	
        J.C.S.~Bacelar\inst{4},
        B.~Bantes\inst{5},
        O.~Bartholomy\inst{2},
        D.E.~Bayadilov\inst{2,3},
        R.~Beck\inst{2},
        Y.A.~Beloglazov\inst{3},
        R.~Castelijns\inst{4},
        V.~Crede\inst{2,6},
        H.~Dutz\inst{5},
        D.~Elsner\inst{5},
        R.~Ewald\inst{5},
	F.~Frommberger\inst{5},
        C.~Funke\inst{2},
        R.~Gregor\inst{7},
        A.B.~Gridnev\inst{3},
        E.~Gutz\inst{2},
	W. Hillert\inst{5},
        S.~H\"offgen\inst{5},
        J.~Junkersfeld\inst{2},
        H.~Kalinowsky\inst{2},
        S.~Kammer\inst{5},
        V.~Kleber\inst{5},
        Frank~Klein\inst{5},
        Friedrich~Klein\inst{5},
        E.~Klempt\inst{2},
        M.~Kotulla\inst{1,7},
        B.~Krusche\inst{1},
        M.~Lang\inst{2},
        H.~L\"ohner\inst{4},
        I.V.~Lopatin\inst{3},
        S.~Lugert\inst{7},
        D.~Menze\inst{5},
        J.G.~Messchendorp\inst{4},
        V.~Metag\inst{7},
        V.A.~Nikonov\inst{2,3},
        M.~Nanova\inst{7},
        D.V.~Novinski\inst{2,3},
        R.~Novotny\inst{7},
        M.~Ostrick\inst{5,11},
        L.M.~Pant\inst{7,12},
        H.~van Pee\inst{2,7},
        M.~Pfeiffer\inst{7},
        A.~Roy\inst{7,13},
        A.V.~Sarantsev\inst{2,3},	
        S.~Schadmand\inst{7,14},
        C.~Schmidt\inst{2},
        H.~Schmieden\inst{5},
        B.~Schoch\inst{5},
        S.V.~Shende\inst{4},
        V.~Sokhoyan\inst{2},
        A.~S{\"u}le\inst{5},
        V.V.~Sumachev\inst{3},
        T.~Szczepanek\inst{2},
        U.~Thoma\inst{2,7},
        D.~Trnka\inst{7},
        R.~Varma\inst{7,13},
        D.~Walther\inst{5},
        \and C. Wendel\inst{2}
\newline(The CBELSA/TAPS collaboration)
\mail{B. Krusche, Klingelbergstrasse 82, CH-4056 Basel, Switzerland,
\email{Bernd.Krusche@unibas.ch}}
}
\institute{Department Physik, Universit\"at Basel, Switzerland
           \and Helmholtz-Institut f\"ur Strahlen- und Kernphysik
                der Universit\"at Bonn, Germany
	   \and Petersburg Nuclear Physics Institute, Gatchina, Russia	
           \and KVI, University of Groningen, The Netherlands
           \and Physikalisches Institut der Universit\"at Bonn, Germany
           \and Department of Physics, Florida State University, Tallahassee,
           USA
           \and II. Physikalisches Institut, Universit\"at Giessen, Germany
	   \and Laboratory of Mathematical Physics, Tomsk Polytechnic University, 634034 Tomsk, Russia
           \and Department of Physics and Astronomy, University of Georgia, Athens, Georgia 30602, USA
	   \and Institut f\"ur Kernphysik, Universit\"at Mainz, 55099 Mainz, Germany
           \and present address: University of Mainz, Germany
           \and on leave from Nucl. Phys. Division, BARC, Mumbai, India
           \and on leave from Department of Physics, Indian Institute of Technology
	        Mumbai, India
	   \and present address: Institut f\"ur Kernphysik, Forschungszentrum J\"ulich, Germany
}
\authorrunning{I. Jaegle et al.}
\titlerunning{Quasi-free photoproduction of $\eta '$-mesons off the deuteron}

\abstract{Quasi-free photoproduction of $\eta '$ mesons off nucleons bound in 
the deuteron has been measured with the combined Crystal Barrel - TAPS
detector. The experiment was done at a tagged photon beam 
of the ELSA electron accelerator in Bonn for incident photon energies from 
the production threshold up to 2.5 GeV. The $\eta '$-mesons have been detected 
in coincidence with recoil protons and recoil neutrons. The quasi-free proton 
data are in good agreement with the results for free protons, indicating
that nuclear effects have no significant impact. The coincidence with
recoil neutrons provides the first data for the $\gamma n \rightarrow n\eta '$ 
reaction. In addition, also first estimates for coherent $\eta '$-production 
off the deuteron have been obtained. In agreement with model predictions,
the total cross section for this channel is found to be very small, at most at 
the level of a few nb.
The data are compared to model calculations taking into account contributions 
from nucleon resonances and $t$-channel exchanges.   
} 
\maketitle

\section{Introduction}
The complex structure of the nucleon is still one of the greatest challenges
for the understanding of the strong interaction in the low energy, 
non-perturbative
regime. One expects that, like in nuclear structure physics, the main 
properties of the interaction are reflected in the excitation spectrum of 
the nucleon, but so far the correspondence between model predictions and 
experimentally observed states is quite unsatisfactory. All constituent 
quark models predict more states than have been observed. This problem of 
`missing resonances' becomes more severe the higher the excitation energy. 
However, the experimental data base is dominated by elastic scattering of 
charged pions off the nucleon, which profits from large hadronic cross 
sections, but is biased against states that couple only weakly to $N\pi$. 
The combination of continuous wave electron accelerators with sophisticated 
4$\pi$ detection systems now allows the study of photon induced reactions with 
at least comparable precision as hadron induced reactions. Therefore, 
photoproduction of mesons has developed into a prime tool for the 
investigation of the nucleon excitation scheme
\cite{Krusche_03,Burkert_04}. 

\begin{figure}[th]
\resizebox{0.5\textwidth}{!}{%
  \includegraphics{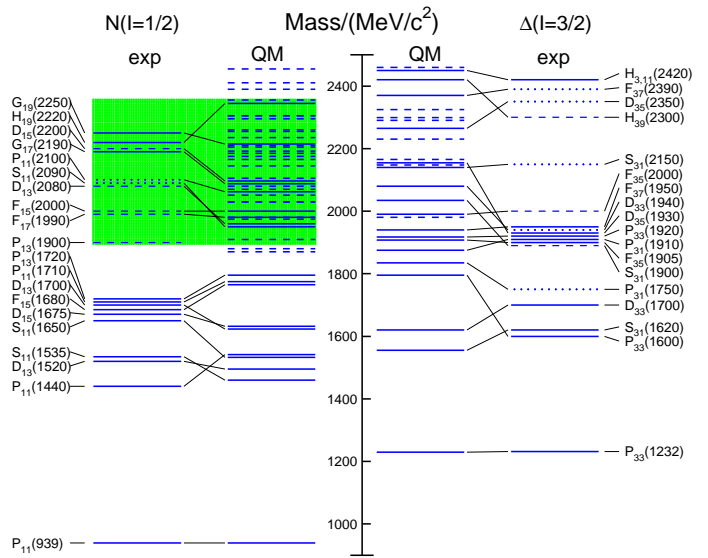}
}
\caption{Comparison of experimentally established nucleon resonances 
\cite{PDG} (left and right) to predictions in the framework of 
non-relativistic quark models \cite{CapRob} (center). 
The shaded (color: green) area indicates the range accessible for $\eta '$
photoproduction in the present experiment.}
\label{fig:levels}       
\end{figure}

Photoproduction of light mesons like pions at high incident photon energies 
involves many partial waves, so that the interpretation of the data requires 
sophisticated partial wave analyses. Such programs are under way and will 
largely profit from the combination of polarized photon beams with polarized 
targets giving access to single and double polarization observables.
However, alternatively due to the suppression of higher partial waves,
the photoproduction of heavier mesons close to their production thresholds 
may give access to resonances which contribute only weakly to other channels. 
Photoproduction of $\eta$-mesons, which is completely dominated in the 
threshold region by the S$_{11}$(1535) resonance, is the best studied example 
for this approach 
\cite{Krusche_95,Krusche_97,Ajaka_98,Bock_98,Armstrong_99,Thompson_01,Renard_02,Dugger_02,Crede_05,Nakabayashi_06,Bartholomy_07,Elsner_07,Denizli_07,Crede_09,Williams_09,Sumihama_09,McNicoll_10}.	
Since the mass of the $\eta '$ ($m_{\eta '}\approx $  958 MeV \cite{PDG}) 
is much higher than the $\eta$-mass ($m_{\eta}\approx $  548 MeV \cite{PDG}), 
resonances contributing to $\eta '$ threshold production may have masses 
around 2 GeV. Of course, lower lying resonances may also contribute
due to their large widths.
Because of their iso-scalar nature, both, $\eta$ and $\eta '$ offer the 
additional selectivity that only $N^{\star}$ resonances can couple to 
$N\eta, N\eta '$. Excited $\Delta$-states can emit these mesons only when 
decaying to other $\Delta$'s, in particular the $\Delta$(1232), and thus 
contribute to the $\eta '\pi$-channel but not to single $\eta '$-production 
(again such processes have been recently intensively studied for 
$\eta$-production in the $\eta\pi^0$-channel
\cite{Nakabayashi_06,Ajaka_08,Gutz_08,Horn_08a,Horn_08b,Kashevarov_09,Gutz_10}).    
Therefore, $\eta '$ threshold production is expected to have a large
sensitivity to $N^{\star}$ resonances at excitation energies, where the 
missing resonance problem is most severe. This is illustrated in Fig. 
\ref{fig:levels}, where the experimentally observed nuclear excitation scheme 
is compared to model predictions. 

Until recently, $\eta '$ photoproduction was not much explored, not even for 
the proton. In an early attempt, Mukho\-pad\-hyay and coworkers \cite{Nimai_95} 
analyzed bubble chamber data with an effective Lagrangian model and concluded 
that the dominant contribution comes from the excitation of a D$_{13}$(2080) 
resonance. Analyses of a more recent measurement with the SAPHIR detector 
\cite{Ploetzke_98,Chiang_03} claimed contributions from different resonances 
(S$_{11}$, P$_{11}$) and strong $t$-channel contributions. However, these
data are not in good agreement with three later measurements with the CLAS 
detector at Jlab \cite{Dugger_06,Williams_09} and the Crystal Barrel/TAPS 
setup at ELSA \cite{Crede_09}. These second generation experiments, which 
profit from much better counting statistics and better control of systematic 
effects due to the use of highly efficient detector systems with large and 
uniform solid angle coverage, clearly supersede the previous data.  
Nakayama and Haberzettl \cite{Nakayama_06} presented an 
analysis of the earlier CLAS-data in the framework of an effective Lagrangian 
model. They found possible contributions from S$_{11}$, P$_{11}$, P$_{13}$, and 
D$_{13}$ resonances. However, also these results are far from being
unique since the available cross section data do not sufficiently constrain 
them. 

Additional information may be obtained by exploring the iso-spin degree of
freedom. The electromagnetic excitations of N$^{\star}$ resonances are 
iso-spin dependent. Resonances which are only weakly excited
for the proton may give stronger signals for the neutron and vice versa. 
Interference patterns between different resonances and between
resonances and background contributions may change due to sign changes of 
the electromagnetic couplings. Again, $\eta$ photoproduction with the 
S$_{11}$(1535) - D$_{13}(1520)$ interference 
\cite{Krusche_95b,Weiss_03,Jaegle_08}
and the prominent excitation structure above the S$_{11}$ range, which is only
seen for the neutron \cite{Jaegle_08,Kuznetsov_07,Miyahara_07}, is a very 
instructive example. 

Due to the technical problems involved in the measurement of small production 
cross sections off bound nucleons, photoproduction of $\eta '$-mesons off 
the neutron had not been studied up to now. Here, we report the first results 
for quasi-free $\eta '$-photoproduction off protons and neutrons bound in 
the deuteron. The paper is organized in the following way. 
Sections \ref{sec:setup} and \ref{sec:analysis} summarize the experimental setup
and the data analysis. The bases of the models the data are compared to are
discussed in section \ref{sec:models}. The results are summarized
in section \ref{sec:results}. First, the data in coincidence with recoil protons 
are compared to free proton data as a check of the quasi-free production 
hypothesis. Both, quasi-free proton and neutron data are then compared to 
model fits. In addition to the quasi-free data, a first estimate for the cross 
section of the coherent process $\gamma d\rightarrow d \eta '$ was extracted. 
Final conclusions are given in section \ref{sec:conclu}.

\section{Experimental setup}
\label{sec:setup}

\begin{figure*}[th]
\resizebox{1.0\textwidth}{!}{%
  \includegraphics{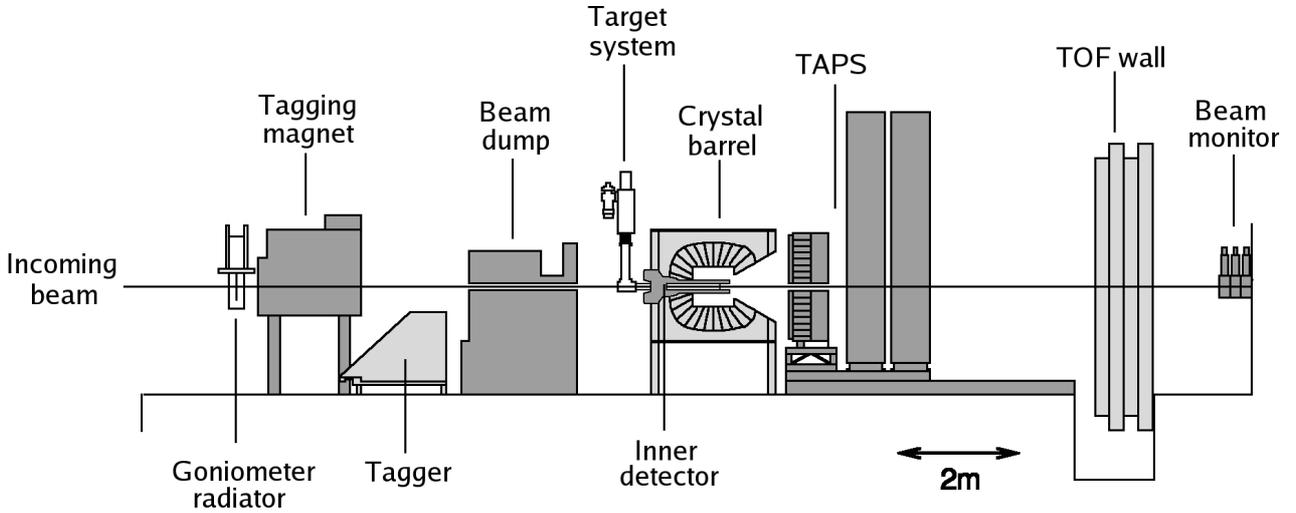}
}
\caption{Overview of the experimental setup at the ELSA accelerator.
}
\label{fig:setup}       
\end{figure*}

\begin{figure}[th]
\resizebox{0.48\textwidth}{!}{%
  \includegraphics{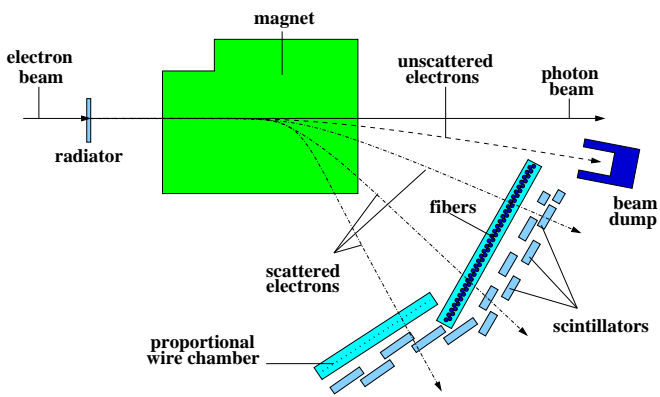}
}
\caption{Setup of the tagging spectrometer.
}
\label{fig:tagger}       
\end{figure}

The experiment was done at the electron stretcher accelerator ELSA in Bonn
\cite{Husmann_88,Hillert_06}. For the measurements discussed here, electron 
beam energies of 2.6~GeV and 3.2~GeV have been used. Real photons were produced
with the brems\-strahlung technique. Their energies were tagged via the
momentum analysis of the scattered electrons by a magnetic spectrometer, which
is schematically shown in Fig. \ref{fig:tagger} (see \cite{Mertens_08} for details).
For this experiment only the part of the focal plane covered by the scintillating 
fiber detector but not the part covered by the proportional wire chamber was used.
This limited the maximum tagged photon energies to 80 \% of the electron beam 
energy. The different beam settings are summarized in Tab. \ref{tab:beam}.

\begin{table}[h]
\begin{center}
\caption{Summary of beam times. $E_{e^-}$: electron beam energy, $E_{\gamma_t}$:
maximum energy of tagged photons, $E_{pol}$: energy of maximum linear photon beam
polarization, $\Phi_o$: energy integrated electron flux. Total life time:
beam time multiplied by acquisition life time.} 
\label{tab:beam}       
\begin{tabular}{|c|c|c|c|c|c|}
\hline\noalign{\smallskip}
characteristics & A & B & C & D & E\\
\hline
$E_{e^-}$ [GeV] & 2.6 & 2.6 & 3.2 & 3.2 & 3.2\\
\hline
$E_{\gamma_t}$ [GeV] & 2.0 & 2.0 & 2.5 & 2.5 & 2.5\\
\hline
$E_{pol}$ [GeV]  & 1.0  & 1.0 & unpol. & 1.2 & 1.6 \\
\hline
total life time [h]  & 138 & 18 & 189 & 25 & 25 \\
\hline
$\Phi_o$ [10$^{7}e^-$/s] & 1.75 & 1.6 & 1.6 & 2.8 & 2.8\\ 
\noalign{\smallskip}\hline
\end{tabular}
\end{center}
\end{table}

Due to these settings and the typical $1/E_{\gamma}$ intensity distribution
of the photon flux, the average time integrated flux at photon energies 
above 2 GeV was roughly lower by a factor of two than the flux at lower energies.
The largest part of the 3.2 GeV beam time was done with a copper radiator 
foil of 0.3 \% radiation lengths thickness, producing unpolarized 
brems\-strahlung. For a small part of this beam time and for the running with
the 2.6 GeV electron energy a diamond radiator was used to produce a linearly
polarized photon beam via coherent bremsstrahlung (see \cite{Elsner_09} 
for details about linear polarization) for the extraction of photon beam
asymmetries. However, the statistical quality of this observable was marginal 
for $\eta '$ production, since for most of the beam time the setting of the 
polarization peak was optimized for $\eta$-production at lower incident photon 
energies (see Tab. \ref{tab:beam}). Therefore photon asymmetries have not been
analyzed for the $\eta '$-channel. 

The target consisted of a vertically mounted cryostat attached to a tube 
entering the Crystal Barrel detector from the upstream side. The target cell 
itself was a capton cylinder (0.125 mm foil thickness) with a diameter of 
3 cm and a length of 5.3 cm, filled with liquid deuterium (surface 
thickness 0.26 nuclei/barn). 
The reaction products emerging from the target were detected with electromagnetic 
calorimeters covering almost the full solid angle; the Crystal Barrel (CB) 
detector (1290 CsI crystals covered the full azimuthal angle for polar angles 
between 30$^{\circ}$ and 168$^{\circ}$) \cite{Aker_92} and the TAPS detector 
(528 BaF$_2$ crystals mounted as hexagonal forward wall covered polar angles down 
to 4.5$^{\circ}$) \cite{Novotny_91,Gabler_94}. Plastic scintillator detectors 
in front of the TAPS modules and a scintillating fiber detector \cite{Suft_05} 
inside the Barrel covering the same polar-angle range were used for charged 
particle identification. A schematic view of the full arrangement is shown in 
Fig. \ref{fig:setup}, more details can be found in \cite{Mertens_08}, where 
apart from the target an identical setup was used. The time-of-flight walls 
were mounted but not used for this experiment.

\begin{figure}[hbt]
\vspace*{0.cm}
\resizebox{0.24\textwidth}{!}{%
  \includegraphics{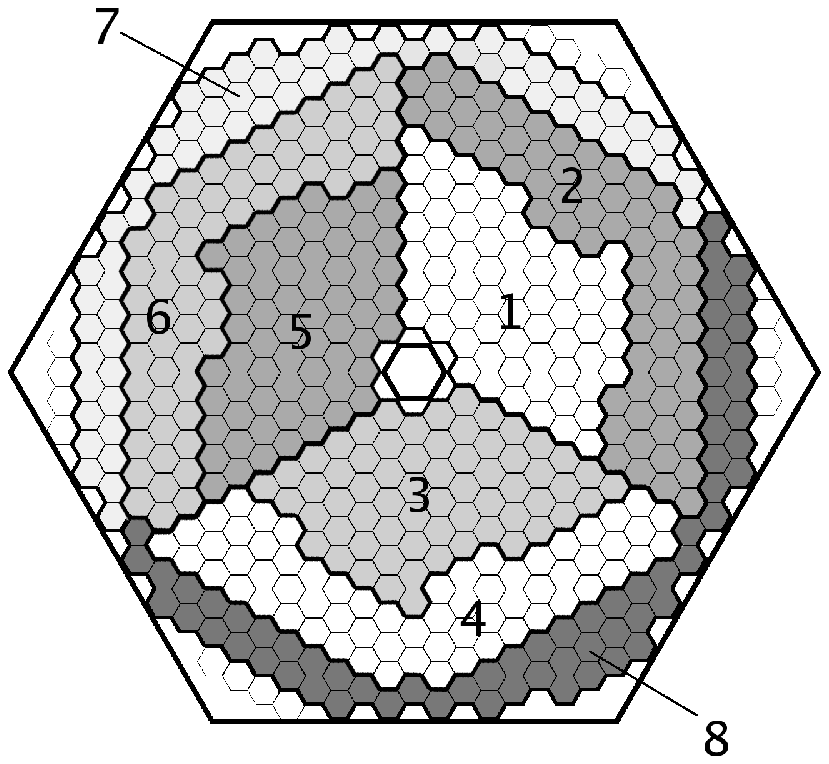}
}
\resizebox{0.24\textwidth}{!}{%
  \includegraphics{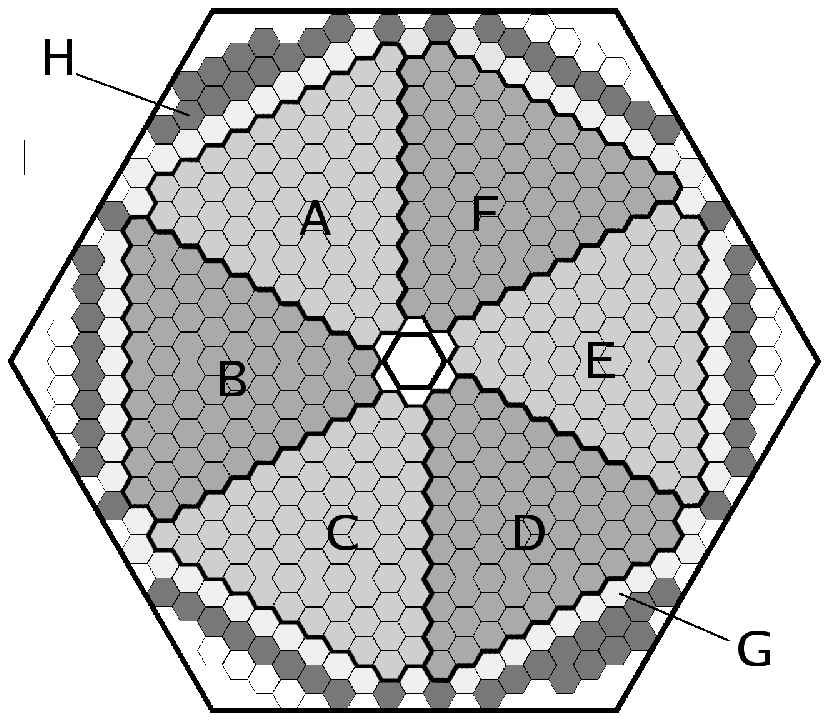}
}
\caption{Front view of the TAPS forward wall. Left hand side: logical
segmentation for the LED-low trigger, right hand side: logical segmentation
for the LED-high trigger (see text).
}
\label{fig:calor}       
\end{figure}

The first-level hardware trigger for the experiment was exclusively based on 
signals from 
the TAPS forward wall detector. The reason is that the CB was read out by 
photodiodes without timing information. Measurements of reactions 
off the free proton can use signals from recoil protons traversing the 
Inner detector. However, in order to have an identical trigger setting for 
quasi-free production off the proton and off the neutron this option was not 
used. The modules of the TAPS detector were equipped with two independent 
leading edge discriminators, combined in two different ways into logical 
groups (see Fig. \ref{fig:calor}). For most of them (ring 12 - 5 from outer 
edge to center) the lower threshold was set to $\approx$55 MeV (LED-low). 
It was set to 80 MeV, 135 MeV, 270 MeV for rings 4, 3, 2, respectively. 
The inner-most ring was not allowed to trigger.
The LED-high thresholds were set to 70 MeV for rings 9 - 7, rising from
105 MeV (ring 6) to 180 MeV (ring 2). Again, the inner-most ring was not
allowed to trigger and the three outer rings (block G) had no LED-high.
The first level trigger included two components: one or more 
LED-low discriminators from at least two logical sections above threshold or 
at least one LED high discriminator above threshold. In the second case, a 
second-level trigger from the FAst Cluster Encoder (FACE) of the Crystal 
Barrel, indicating at least two separated hits in the Barrel, was required 
in addition. All first level triggers thus required detection of at least one 
photon in TAPS. Such a trigger is only efficient for reactions with a high 
multiplicity of photons like the 
$\eta '\rightarrow \pi^0\pi^0\eta\rightarrow 6\gamma$ 
or the $\eta\rightarrow 3\pi^0\rightarrow 6\gamma$ decays. 
But even then the trigger efficiency for mesons at backward angles is not 
large. In principle, also events where a recoil nucleon is detected in TAPS
might activate the hardware trigger. This would, however, lead to uncontrollable
trigger efficiencies since the LED thresholds are only calibrated for photons
(recoil nucleons have different signal shapes in BaF$_2$ scintillators) and
the energy deposited by neutrons is more or less random. Therefore only events
where photons alone (identified by non-firing veto detectors and time-of-flight 
versus energy) fulfilled the first level trigger conditions were accepted in
the analysis.

\section{Data Analysis}
\label{sec:analysis}
\subsection{Particle and reaction identification and extraction of cross 
sections}
\label{subsec:analysis}
Photoproduction of $\eta '$-mesons was identified via the 
$\eta '\rightarrow\eta\pi^0\pi^0\rightarrow 6\gamma$ decay chain, which has a
branching ratio of 8\%. Cross sections were extracted for four different
reaction types. The two most important ones are quasi-free production off the
proton $\gamma d\rightarrow (n)p\eta '$, which requires coincident detection
of an $\eta '$ and a recoil proton and quasi-free production off the neutron
$\gamma d\rightarrow n(p)\eta '$ via detection of the $\eta '$ together with
a recoil neutron. For the control of systematic uncertainties (see below) also
the inclusive reaction $\gamma d\rightarrow (np)\eta '$ with no condition
for recoil nucleons was analyzed. Finally, also an estimate of the fully
inclusive reaction $\gamma d\rightarrow X\eta '$ was obtained, where also
final states like $\eta '\pi$ contribute.

\begin{figure}[tht]
\resizebox{0.5\textwidth}{!}{%
  \includegraphics{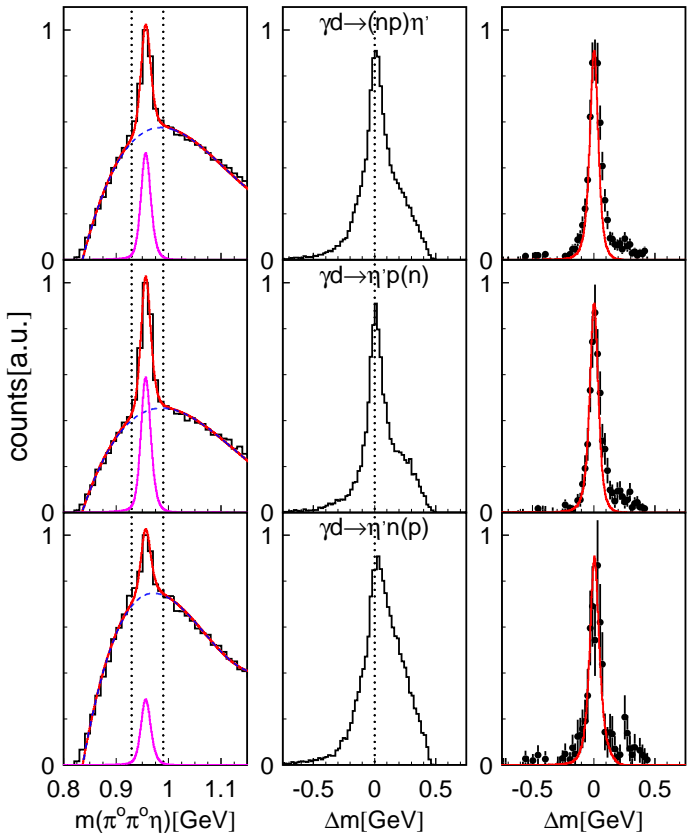}
}
\caption{Top to bottom: reactions $\gamma d\rightarrow (np) \eta '$,
$\gamma d\rightarrow p(n) \eta '$, $\gamma d\rightarrow n(p) \eta '$.
Left hand side: 6$\gamma$ invariant mass spectra, (Blue) dashed \
curves: background fit, (red) solid histograms: sum of background and 
$\eta '$-peaks, in addition: simulated line-shapes (pink). 
Center: missing mass spectra for cut on $\eta '$ invariant mass peaks, 
right hand side: (black) points: missing mass spectra for events in 
invariant mass peak after background subtraction, solid (red) curves: 
simulated line-shapes. All spectra for incident photon energies from
threshold to 2 GeV (integrated over all beam times and the full
polar angle range).
}
\label{fig:iden}       
\end{figure}

In the first step of the analysis photon and neutron
candidates (called `neutral hits') were separated from proton candidates 
(called `charged hits'). This was done in the CB with the help of the 
scintillating fiber detector and in TAPS with the charged particle veto 
detectors. In TAPS a hit was assigned to `charged' if the veto of any cluster 
module or the veto of any neighbor of the central module of the cluster had 
responded (even if the neighbor module itself had no signal above threshold). 
The latter condition applies to charged particles which traverse the edge of 
a veto but deposit their energy in the neighbor module (due to large impact angles). 
All other hits were assigned to `neutral'.
In the Barrel, a hit was assigned to `charged', if at least two layers of the
Inner-detector had recorded a hit within $\pm 10^{\circ}$ of azimuthal angle.
It was assigned to `neutral' if no layer had fired within this azimuthal angle.
Hits with one responding layer of the Inner-detector were discarded. 
In the TAPS forward wall, correct identification of protons, neutrons, and 
photons can be additionally controlled with a time-of-flight versus energy 
analysis, while in the CB no direct separation of photons and neutrons 
is possible (more details on particle identification are given in 
\cite{Mertens_08} (CB) and \cite{Bloch_07} (TAPS)). 

In the next step, events with at least six `neutral' hits as candidates 
for the $\eta '$-decay photons were selected and assigned to four partly 
overlapping classes corresponding to the above defined reaction types: 
six `neutral' and one `charged' hit for the $(n)p\eta '$ final state,
seven `neutral' for the $n(p)\eta '$ final state, six or seven neutral
or six neutral and one charged for $(np)\eta '$, and at least six `neutral' 
without any further condition for $X\eta '$. 

The identification of the $\eta 'N$ final states was then based on 
a combined invariant and missing mass analysis. The invariant mass analysis 
identified the $\eta '$, the missing mass analysis excluded events where
further mesons have been produced but have escaped detection (except for 
the $X\eta '$ final state where such events were included). 

The invariant mass of all possible disjunct photon pairs was calculated. 
Only events having at least one combination of six `neutral'
hits to two photon pairs with invariant masses between 110 and 160 MeV 
(pions) and one pair between 500 and 600 MeV ($\eta$) were kept. In cases 
where the photons could be combined in more than one way to fulfill this 
condition, the `best' combination was chosen via a $\chi^2$ minimization: 
\begin{equation}
\chi^2 = \sum_{k=1}^{3}\frac{(m_{k}(\gamma\gamma)-M_k)^2}{(\Delta
m_{k}(\gamma\gamma))^2}
\end{equation}
where for each disjunct combination of the six photons into three pairs
the invariant masses are ordered so that 
$m_{1}(\gamma\gamma)\leq m_{2}(\gamma\gamma)\leq m_{3}(\gamma\gamma)$.
The $\Delta m_{k}(\gamma\gamma)$ are their uncertainties (computed
event-by-event from the detector resolution for energies and angles)
and $M_{k}$ is the $\pi^0$-mass for $k=1,2$ and the $\eta$-mass for $k=3$. 
The above case applies to events with exactly six `neutral' hits, where in 
total 15 different combinations are possible (events with recoil proton or 
without detected recoil nucleon). For candidates for the quasi-free reaction 
off the neutron (seven `neutral' hits) one must in addition loop over the 
unpaired hit, since in CB photons and neutrons cannot be distinguished. 
This corresponds at maximum to 105 combinations, giving rise to larger 
combinatorial background. In this case, the hit which was not assigned to a
pion- or $\eta$-decay photon is accepted as neutron candidate. 
In the case of the $X\eta '$ final state even higher multiplicities may 
occur.   

As in \cite{Mertens_08} the nominal masses of the mesons 
were then used as a constraint to improve the experimental resolution by 
re-calculating the measured photon energies from
\begin{equation}
E'_{1,2}=E_{1,2}\frac{m_{\pi^0, \eta}}{m_{\gamma\gamma}} \ ,
\end{equation}   
where $E_{1,2}$ are the  measured photon energies, $E'_{1,2}$ the re-calculated,
$m_{\pi^0,\eta}$ are the nominal $\pi^0, \eta$ masses, and $m_{\gamma\gamma}$ 
the measured invariant masses. 

The obtained 6-photon invariant mass distributions using the re-calculated
$\gamma$-energies are shown in Fig. \ref{fig:iden}, (left column) for the 
inclusive reaction $(np)\eta '$ and in coincidence with recoil protons 
$(n)p\eta '$ and neutrons $(p)n\eta '$. In all cases a clear peak is visible 
at the nominal $\eta '$-mass of 958 MeV. The shape of the invariant mass peaks 
has been generated via a Monte Carlo simulation with the GEANT package 
\cite{Brun_86} and fitted to the data together with a background polynomial.
The peak-to-background is best for $(n)p\eta '$, intermediate for $(np)\eta '$,
and worst for $(p)n\eta '$. This is as expected from the above discussion:
events with seven `neutral' hits have a much larger chance for combinatorial
background (for example from $\eta\pi^0$, $\eta '\pi^0$, $\eta n$ final states,
when a photon is falsely assigned as neutron or vice versa) 
than events with six `neutral' and one `charged' hit. This is also reflected
in the background of the missing mass spectra, which is much more pronounced 
when the invariant mass background is not subtracted (compare center and right
column in Fig. \ref{fig:iden}). 

For the missing mass analysis the recoil nucleons were treated as missing 
particles, no matter if they were detected or not.
The missing mass $\Delta m$ of the reaction was calculated for quasi-free 
production of $\eta '$ mesons off nucleons via:
\begin{equation}
\Delta m = \left|{\mbox{\bf P}_{\gamma}+\mbox{\bf P}_{N}-\mbox{\bf P}_{\eta '}}\right|
-m_{N}   \ ,
\end{equation}
where ${\mbox{\bf P}_{\gamma}}$, ${\mbox{\bf P}_{N}}$, ${\mbox{\bf P}_{\eta '}}$
are the four-momenta of the incident photon, the incident nucleon (assumed to 
be at rest), and the produced $\eta '$-meson; $m_N$ is the nucleon mass. 
The resulting distributions peak around zero for quasi-free $\eta '$ production 
and are somewhat broadened by the momentum distribution of the bound nucleons, 
which was neglected. The distributions are shown in Fig. \ref{fig:iden}, 
center column.
They have been constructed for an invariant mass window from 930 - 990 MeV
(see Fig. \ref{fig:iden}, left column). Since the background cannot be 
completely removed by cuts on invariant mass and missing mass, in the final 
step the invariant mass spectra have been analyzed (i.e. fitted by line shape 
and background) for each bin of missing mass. The resulting missing mass 
spectra corresponding to the invariant mass peaks are shown in 
Fig. \ref{fig:iden}, right hand side. Background is much reduced and the shapes 
of the missing mass peaks are quite well reproduced by the Monte Carlo 
simulation. 
The residual background at positive missing masses is mainly
due to the $\eta '\pi$ final state, from events where the pion escaped detection.
It becomes more important at incident photon energies above 1.6 GeV
(see also Fig. \ref{fig:total_incl}) and extends into the range of the missing 
mass peak. This explains also the deviation of the simulated line shape from
the data at positive missing mass.
Therefore at energies above 1.6 GeV only events with missing mass 
between -200 and 0 MeV were accepted. This reduces counting statistics by 
a factor of two but guarantees negligible background contamination. 
The analysis described above was done independently for each bin of incident 
photon energy and $\eta '$ polar angle.

\begin{figure}[ht]
\resizebox{0.48\textwidth}{!}{%
  \includegraphics{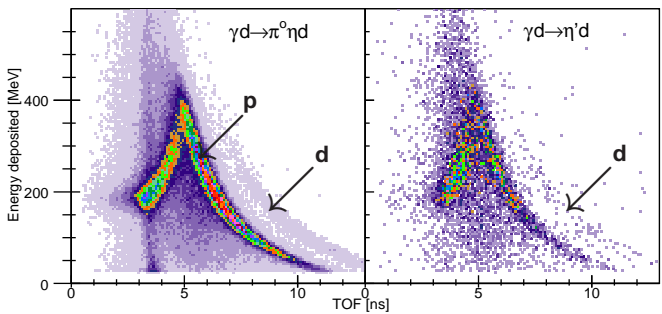}
}
\caption{Time-of-flight versus energy for charged particles in TAPS under the
condition of invariant mass signals (cuts on invariant mass peaks, no
background subtraction) for the $\pi^0\eta$ channel (left hand side),
and the $\eta '$ channel (right hand side). The prominent band corresponds to 
protons, the less prominent band to deuterons.
}
\label{fig:tofe}       
\end{figure}

\begin{figure}[ht]
\resizebox{0.5\textwidth}{!}{%
  \includegraphics{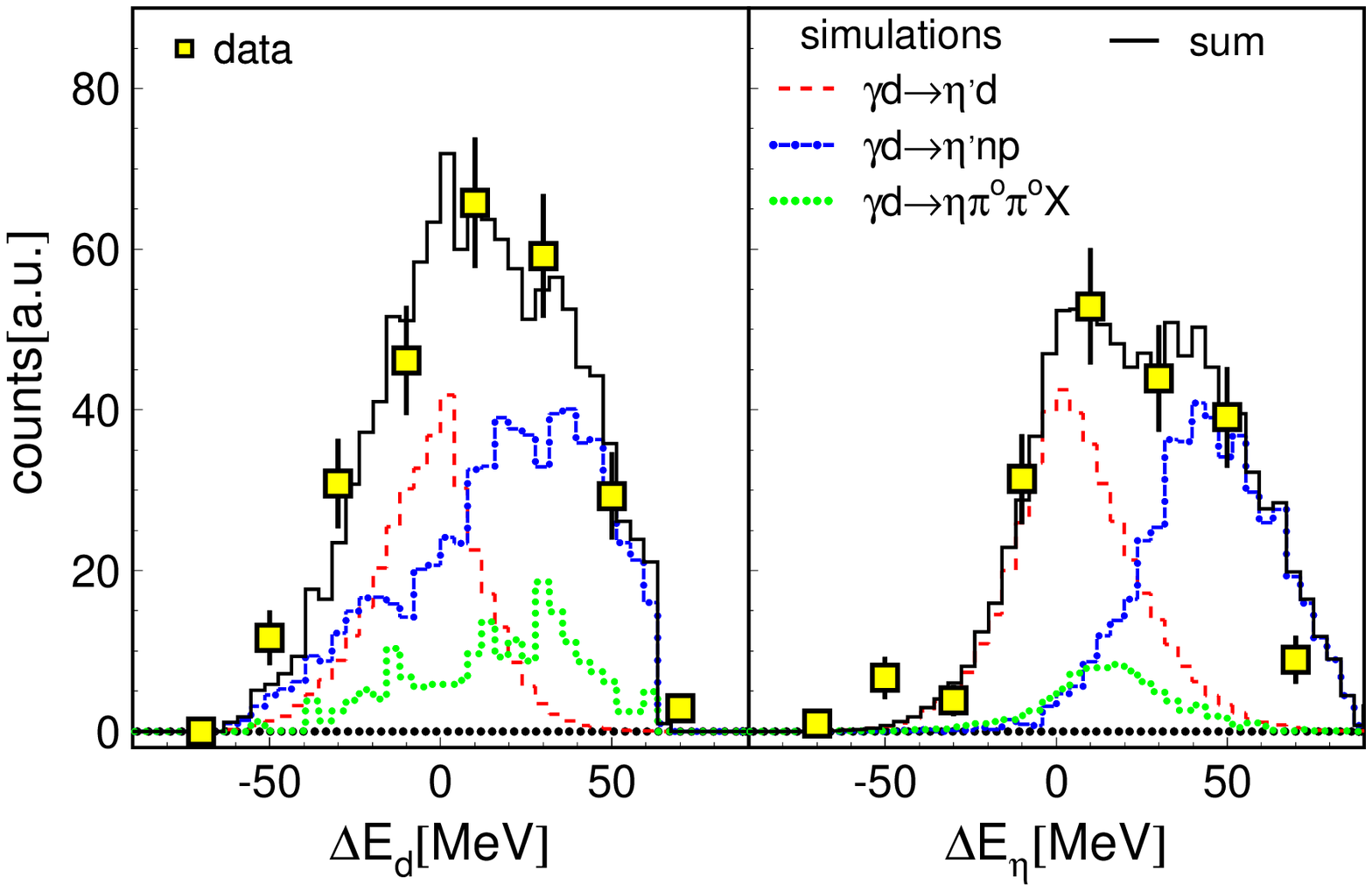}
}
\resizebox{0.5\textwidth}{!}{%
  \includegraphics{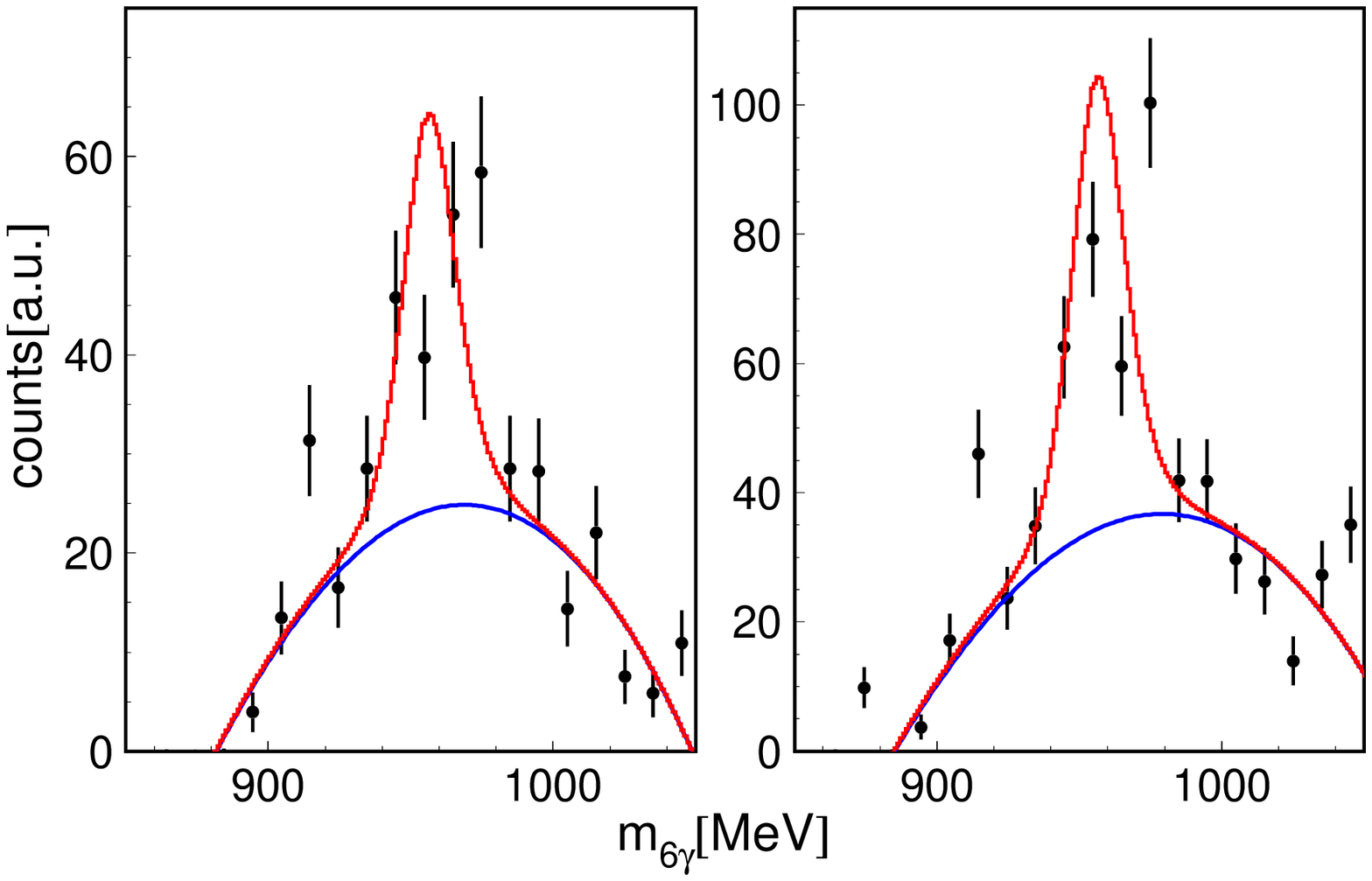}
}
\caption{Upper part: missing energy. Left hand side: deuteron missing energy, 
Right hand side: $\eta '$ missing energy after cut on deuteron missing energy 
between $\pm$35 MeV. Points: data, histograms: simulation of signal and background 
shapes.
Bottom part: invariant mass spectra. Cut on deuteron missing energy $\pm$35 MeV,
cut on $\eta '$ missing energy: $\pm$25 MeV (left hand side), 
$\pm$50MeV (right hand side). Points: data, fitted with background polynomial 
plus simulated line shape.
}
\label{fig:iden_coh}       
\end{figure}

Absolute cross sections were derived from the target density, the incident 
photon flux, the decay branching ratio, the detection efficiency for the 
$\eta '\rightarrow 6\gamma$ decay, and the detection efficiency for recoil 
nucleons. The detection efficiency was determined via Monte Carlo simulations 
using the GEANT3 package \cite{Brun_86}, which included all features of the
detector system and all software cuts for particle and reaction identification.
Events for the final states $(np)\eta '$, $(p)n\eta '$, and $(n)p\eta '$
were generated evenly distributed in phase-space including the effects of
nuclear Fermi motion, tracked with the GEANT package, and analyzed in the 
same way as the experimental data. The tracking of recoil nucleons was done
with the GEANT-CALOR program package \cite{Zeitnitz_01}, which is optimized
for hadronic interactions from the few MeV to several GeV range, including
the interactions of low energy neutrons. The efficiency correction was then
done in the usual way as function of incident photon energy and meson cm polar 
angle.

Finally, an estimate of the cross section for the coherent reaction 
$\gamma d\rightarrow \eta ' d$ was extracted in the following way. 
Deuterons in the TAPS detector can be identified via a time-of-flight versus 
energy analysis (time-of-flight path 1.18 m) as shown in Fig. \ref{fig:tofe}, 
where only charged hits (identified by the veto detectors) are included. 
The distribution at the right hand side of Fig. \ref{fig:tofe} is obtained for 
6-photon events with an invariant mass corresponding to the $\eta '$-meson. 
The left hand side of the figure shows for comparison the result for the 
$\eta\pi^0$ channel, which shows a much more pronounced deuteron band since 
it has a larger coherent component. Events can then be selected by a cut on 
the deuteron band, which gives a quite clean data sample for the 
$d\eta '$ final state. 

However, due to kinematic reasons (mesons from the coherent reaction are 
mostly emitted to forward angles due to the nuclear form factor) most 
deuterons are emitted into the solid angle of the Barrel, where they cannot 
be identified via time-of-flight. Therefore a more complicated analysis based 
on reaction kinematics was necessary. The major steps are summarized in 
Fig. \ref{fig:iden_coh}. For deuterons in the Crystal Barrel, first the Inner 
detector was used for charged particle identification and it was required that 
the deuteron candidate was co-planar with the $\eta '$-meson. Further 
exploiting the two-body kinematics of the final state, the cm-energies 
(photon - deuteron center of momentum system) derived from the final state 
four-vectors were compared to the respective values calculated from the 
incident photon energy for the deuteron (deuteron missing energy) and for 
the $\eta '$ ($\eta '$ missing energy). These spectra are compared in 
Fig. \ref{fig:iden_coh}, upper part to Monte Carlo simulations of the coherent 
and breakup process and of residual background from $\eta\pi^0\pi^0$ phase 
space contributions not related to $\eta '$ production. The corresponding 
invariant mass spectra after more or less stringent cuts on missing energy 
are shown in the bottom part of Fig. \ref{fig:iden_coh}. The final signal 
strength was then extracted via a fit of a background polynomial and the 
simulated invariant mass line shape to the data. The systematic uncertainty 
of this procedure was estimated by a variation of the $\eta '$ missing energy 
cut ($\pm$50 MeV, $\pm$25 MeV, -50 MeV - 0 MeV). The first cut cannot 
completely suppress incoherent background, results from the most stringent, 
asymmetrical cut are unfortunately limited by the statistical quality of 
the data.   

\subsection{Systematic uncertainties}

Due to the small cross section of $\eta '$-production, the requirement
of coincident detection of the mesons with recoil particles, and the
low trigger efficiency (only trigger signals from photons in TAPS)
counting statistics were low. Therefore statistical uncertainties 
were rather large (on the order of 15 - 35 \% for the $(p)n\eta '$ final 
state), which makes it difficult to investigate systematic effects
hidden in statistical fluctuations. Therefore, systematic
uncertainties were partly extracted from other reaction channels, 
in particular from the $\eta\rightarrow 6\gamma$ channel.   

Three different types of systematic uncertainties may effect the quasi-free
cross sections: overall uncertainties which cancel exactly in the comparison 
of proton and neutron final state, uncertainties with similar effects, which
cancel to a large extend in ratios, and uncertainties related to specific
final states which do not cancel. 

Into the first category fall the systematic uncertainty of the incident photon
flux, the uncertainty in target thickness (as well as effects from possible
slight misplacements of the target), and the uncertainty from the $\eta '$
decay branching ratios. An estimate of the flux uncertainty was obtained
by a comparison of $\eta$-photoproduction \cite{Jaegle_08} cross sections
obtained by a separate analysis of beam times (A) and (C) 
(see Tab. \ref{tab:beam}). These two beam times used different incident electron
beam energies, so that the same photon beam energies were mapped to different
sections on the focal plane detector. Furthermore, since (C) was using an
unpolarized beam and (A) linear polarization, also the energy dependence of 
the flux was different due to the coherent peak at roughly 1 GeV. Typical
deviations between the $\eta$ cross sections produced from these two beam times
are at the 5 \% level, maximum deviations around 10 \%. Since the results
from both beam times (with approximately equal statistical weight) were
averaged, we estimate the systematic flux uncertainty at 10 \%.
The overall systematic uncertainty coming from target thickness and positioning
is on the order of a few per cent. The systematic uncertainty of the branching
ratios for  $\eta '\rightarrow \eta\pi^0\pi^0\rightarrow 6\gamma$ is around
7 \%. Allowing for some cancellation, we estimate a total normalization 
uncertainty of $\approx 15$ \%. A comparison of the results for
quasi-free production off the proton to free proton results for both the 
$\eta$ and the $\eta '$ channel (see below) did not reveal discrepancies 
beyond this level. 

The second class of systematic uncertainties is dominated by the uncertainty
of the $\eta '$ identification by the missing mass and invariant mass analysis
and the simulation of the $\eta '$ detection efficiency, which are of course
related (the better the respective cuts are reflected by the simulation the
smaller the uncertainty). The simulation of the detection efficiency of photons 
followed by an invariant mass analysis for meson identification is very well
under control for the CB/TAPS setup. This has been tested for example via
a comparison of the results for the $\gamma p\rightarrow p\eta$ reaction
obtained from the analysis of the $\eta\rightarrow 2\gamma$ and
$\eta\rightarrow 6\gamma$ decay channels \cite{Crede_05,Bartholomy_07,Crede_09}.
Agreement with the PDG value of the 
$\Gamma_{\eta\rightarrow 3\pi^0}/\Gamma_{\eta\rightarrow 2\gamma}$ ratio
is reported in \cite{Bartholomy_07} within an uncertainty at the 2 \% level.
Since errors for the $\eta\rightarrow 2\gamma$ channel in photon detection or
invariant mass analysis enter cubed into the $\eta\rightarrow 3\pi^0$ channel,
this sets stringent limits on the uncertainty. Already a 2 \% error
in photon detection efficiency would result in an 8 \% deviation between
the two decay channels.

A further uncertainty is related to the choice of the event generator
used for the simulation. Events were generated evenly distributed in 
phase-space, where the effects of Fermi motion were modeled using the deuteron
wave function in momentum space from \cite{Lacombe_81}. 
Since the correction was done as function of incident photon energy 
and cm polar angle of the $\eta '$, the angular distribution of the $\eta '$
mesons itself is not critical. However, deviations could arise
if for example final state interaction effects modify the correlation
between meson polar angle and kinetic energy or between meson polar angle
and the kinematic variables of the recoil nucleon.
Possible systematic effects of this kind were investigated with a different 
simulation, where the detection efficiency for $\eta '$-mesons 
$\epsilon(T_{\eta '},\Theta_{\eta '})$ and the 
detection efficiency for recoil nucleons $\epsilon(T_{N},\Theta_{N})$, N=n,p
was quasi-factorized and parameterized in dependence of laboratory kinetic 
energies $T_{\eta '}$, T$_{N}$ and polar angles $\Theta_{\eta '}$,
$\Theta_N$ of the particles. Typical efficiencies are
10~\% for $\eta '$ detection (including trigger efficiency), 95 \% for
protons, and 10 - 30~\% for neutrons (depending on energy and including
the identification of the neutron).
These kinematic observables can be directly extracted from the measured
data (the neutron kinetic energy is extracted via the over determined reaction
kinematic from the incident photon energy, the measured $\eta '$ four-vector
and the measured neutron angles). Therefore, an event-by-event efficiency 
correction with the product 
$\epsilon(T_{\eta '},\Theta_{\eta '})\cdot\epsilon(T_{N},\Theta_{N})$
becomes possible, which does not rely on any model assumptions about the
kinematic final state variables.
This efficiency correction does, however, not include the missing mass cut,
which depends on incident photon energy.
A correction for this effect was extracted from a phase-space simulation. 
It does not much depend on details of the event generator, since it uses
only the ratio of two different analyses (with and without missing mass cut)
of the same simulation. Actually as expected from Fig. \ref{fig:iden}
the correction factor is close to 2 for a cut from -200 MeV to 0 (left half of
the peak). 
The results from the two different detection efficiency simulations agreed to
better than 5 \% for all investigated reaction channels, and we assume a
systematic uncertainty in this range.

The effects from the background under the invariant mass peaks (see Fig.
\ref{fig:iden}), which is more important for the neutron channel, have not
been treated as independent systematic uncertainties, they have been included
via the peak - background separation into the statistical uncertainties.
An additional systematic uncertainty could arise from the missing mass
analysis. Variations of the accepted missing mass range show significant
influence on the extracted cross section. This, however, does not seem to
be a problem of the agreement between simulation and data, since the shapes
of the signals agree well at the left hand side of the peaks, but start to
disagree at the right hand side where background from $\eta'\pi$ final states 
is expected. Therefore, for incident photon energies above 1.6 GeV only events 
with $\Delta m<$0 have been accepted.
However, we assign an additional uncertainty rising from 3~\% at threshold 
($\eta\pi$ background starts to contribute above 1.6 GeV) to 10~\% at the 
maximum energy. 
Altogether, independent on the reaction channel, we estimate an uncertainty 
of 6~\% close to threshold up to 12~\% at the highest incident photon energies,
(not including effects of recoil nucleon detection).   

The last class of uncertainties are those related to the detection of recoil 
nucleons, which will not cancel in the comparison of neutron - proton cross 
section ratios. The detection of the recoil nucleons was included in the
simulations using the GCALOR package \cite{Zeitnitz_01}, which is optimized 
for this purpose. For the proton, the quality of this simulation could be 
cross checked with experimental data for the $\gamma p\rightarrow p\eta$
and $\gamma p\rightarrow p\pi^0\pi^0$ reactions which have been measured
with the same setup. The proton detection efficiency was simply determined 
as ratio of the number of events with detected recoil proton to the total 
number of events from these reactions. The efficiencies have then also been
simulated and the simulated and measured values agree within 8 \% for slow 
protons and 4 \% for fast protons. 
Combining all uncertainties except the overall normalization we estimate for 
the quasi-free proton channel 10~\% at threshold rising to 15~\% at 2.5 GeV.

For the neutron detection efficiency there are no direct measurements
with the combined TAPS/CB setup in Bonn. For the TAPS detector it had been
experimentally determined from the $\gamma p\rightarrow n\pi^0\pi^+$
reaction at the MAMI accelerator in Mainz \cite{Hejny_99}. The results
are consistent with the GCALOR simulation, when the conditions of the Mainz 
setup are used ($T_n$=250 MeV: simulated 18.5~\%, from data 19.1 \%).  
The neutron detection efficiency of the CB was measured at the LEAR ring at 
CERN \cite{Schaefer_93}. Results from the present GCALOR simulation
are in good agreement with the LEAR result except for slow neutrons
($T_n<$ 75 MeV), where the efficiency is very dependent on detector thresholds
and the neutron kinetic energy. However, in any case it is necessary to 
determine  `effective' neutron efficiencies which take into account the 
identification of the neutrons out of at least seven neutral hits via the 
invariant mass analysis discussed in Sec. \ref{subsec:analysis}. This could 
only be done by simulations. The reduction of the efficiency under this 
conditions compared to the situation were only neutrons are simulated is 
substantial, of the order of 25~\% - 35~\%. We therefore estimate the absolute
systematic uncertainty for neutron detection at the 15 \% level. Altogether 
a systematic uncertainty of 16~\% at reaction threshold up to 20~\% at 
highest incident photon energies is estimated for the quasi-free neutron 
channel (excluding the overall normalization uncertainty). 

A further systematic uncertainty could arise from the misidentification of 
recoil nucleons. While the loss of events is included in the simulated 
efficiencies, misidentified protons might contaminate the neutron sample or 
vice versa, where the first problem is more severe, due to the smaller 
absolute detection efficiency for neutrons. The properties of the Inner 
detector for proton identification have been studied in detail in
\cite{Suft_05} with simulations and data from the reaction 
$\gamma p\rightarrow \pi^0 p$. The main result was that the 
average efficiency for proton detection (somewhat angle dependent) is 98.9 \% 
(simulation) respectively 98.4 \% (data). Also determined where the 
efficiencies of all three layers of the detector (from data: 94.8 \%
(inner layer), 92.9~\% (middle layer), 88.1 \% (outer layer)). Since in this
experiment the condition for neutrons was that no layer had responded, only
about 0.04 \% of protons may be misidentified as neutrons. The probability that 
a neutron activates a coincidence of two layers (condition for proton) is also
negligible. The TAPS veto detectors have on average an inefficiency for
proton detection at the 4 \% level, depending on kinetic energy. However,
for TAPS additional separation of proton and neutron hits is provided by the 
time-of-flight versus energy analysis. No trace of the typical proton band
was seen for `neutral' events and the possible contamination of the neutron 
sample with protons could be estimated at the 1 \% level. Cross contamination
of the recoil nucleon samples was therefore negligible.

The different systematic uncertainties are summarized in Tab. \ref{tab:systematics}.
It should be noted that the comparison of the quasi-free proton data to free
proton data, as well as the comparison of the two different neutron analyses
(see below) indicate that these estimates are pessimistic.

\begin{table}[h]
\begin{center}
\caption{Summary of systematic uncertainties for the quasi-free reactions.
$^{1)}$ photon flux, target thickness,
decay branching ratios, $^{2)}$ trigger efficiency, $\eta '$ analysis
cuts, $\eta '$ detection efficiency. When two numbers are given first corresponds
to threshold, second to $E_{\gamma}$= 2.5 GeV, and linear interpolation} 
\label{tab:systematics}       
\begin{tabular}{|c|c|c|}
\hline\noalign{\smallskip}
source & $\gamma d\rightarrow (n)p\eta '$  & $\gamma d\rightarrow (p)n\eta '$ \\
\hline
overall normalization$^{1)}$ & 15~\% & 15 \% \\
\hline
\hline
$\eta '$ detection$^{2)}$ & 6~\% - 12~\%  & 6~\% - 12~\%\\
\hline
recoil nucleon detection  & 8~\% - 4~\%  & 15~\%  \\
\hline
total except overall norm.  & 10~\% - 15~\% & 16~\% - 20~\%  \\
\noalign{\smallskip}\hline
\end{tabular}
\end{center}
\end{table}

For the comparison of the quasi-free $\gamma p\rightarrow \eta ' p$ and
the $\gamma n\rightarrow \eta ' n$ reactions, systematic uncertainties 
except the ones from the recoil nucleon detection cancel. However, these 
effects can be controlled in an independent way. 
As discussed above, the cross section is constructed for $\eta '$ mesons in 
coincidence with recoil protons ($\sigma_p$), for $\eta '$ mesons in 
coincidence with recoil neutrons ($\sigma_n$), and for $\eta '$ mesons without 
any condition for recoil nucleons ($\sigma_{np}$). Since coherent production 
processes are very small (see below), the cross sections must be related by 
$\sigma_{np} =\sigma_n + \sigma_p$. Therefore, the neutron cross section can 
be extracted in two independent ways as $\sigma_n$ or as 
$\sigma_{np}-\sigma_p$, one depending only on neutron detection efficiency, 
the other depending only on proton detection efficiency. This method has been
previously tested for $\eta$-photoproduction \cite{Jaegle_08}, where excellent
agreement between the two results was found. Also for $\eta '$ the two 
results are in good agreement, their weighted average $\langle\sigma_n \rangle$
is given as final result for the neutron cross section and the differences 
between them (shown in Figs. \ref{fig:total},\ref{fig:coeff}) are an 
independent estimate of the uncertainties introduced by the recoil particle 
detection.

\section{Reaction Models}
\label{sec:models}
In the absence of any data for polarization observables,
only a preliminary interpretation of the data in the framework of reaction
models, using model dependent constraints, is possible. In 2003 Chiang et al. 
\cite{Chiang_03} have developed a reggeized model for $\eta$ and $\eta '$ 
production ($\eta '$-MAID), which they used to analyze the then available 
proton data. 
They parameterized contributions from nucleon resonances in the usual way in 
terms of Breit-Wigner curves with energy dependent widths. Non-resonant 
Born-terms were neglected since they were expected to be small at not too 
high photon energies due to the small $\eta '$-nucleon-nucleon coupling 
constant $g_{\eta ' NN}$. However, they are included in the most recent 
version of the model used to fit the present data. Contributions from 
$t$-channel vector meson exchange are important and were incorporated via 
Regge trajectories.
They found, that already a model including just one S$_{11}$ resonance 
($W$=1960 MeV) together with the Regge trajectories could reproduce
the available angular distributions for $\gamma p\rightarrow p\eta '$.
Some improvement of the fit was possible by addition of a P-wave resonance,
where a P$_{11}$ or a P$_{13}$, both around $W$ = 1950 MeV, gave equally good
results. However, in the meantime, the database for the proton has been 
much improved and, neither the absolute magnitude nor the extreme 
forward - backward asymmetry of the early angular distributions 
\cite{Ploetzke_98} have been supported by the later experiments 
\cite{Dugger_06,Williams_09,Crede_09}, so that these fits needed to be updated.
For this purpose, the model was extended by addition of a D$_{13}$ resonance. 
It was then fitted simultaneously to the free proton data from CLAS and ELSA, 
to the present quasi-free proton data, and to the quasi-free neutron data. 
The effects from Fermi-smearing, although not important, were taken into 
account for the quasi-free data sets by folding the model results with the 
nucleon momentum distributions. 

In a different approach Nakayama and Haberzettl \cite{Nakayama_06} have 
analyzed the first CLAS-data \cite{Dugger_06}. They extended their relativistic
meson-exchange model \cite{Nakayama_04} for application to the 
$\gamma p\rightarrow p\eta '$ reaction by introducing contributions from
spin-3/2 resonances (the earlier version considered only spin 1/2-states) 
and including energy-dependent resonance widths. In addition to the resonance 
contributions nucleonic s- and u-channel diagrams and, more important, 
mesonic $t$-channel contributions ($\rho$, $\omega$ exchange) are considered. 
However, due to the lack of polarization observables, they find also
different solutions. 
In Ref. \cite{Nakayama_06} also the $\rho$- and $\omega$-Regge trajectories 
have been considered instead of the t-channel rho and omega meson exchanges 
to describe the CLAS data \cite{Dugger_06}. There, it is found that the Regge 
description yields similar results as the t-channel meson exchange. Therefore, 
in the present work, we confine ourselves to their model with t-channel meson 
exchange.
The `minimum' solution with the smallest number of resonances in addition to 
the nucleonic and mesonic currents that gives an acceptable fit 
includes an S$_{11}$(1958) and P$_{11}$(2104) as well as sub-threshold 
P$_{13}$(1885) and D$_{13}$(1823) states (which can be considered as 
non-resonant backgrounds). 
In the following, we call this solution (I), whose parameter values are 
summarized in Table I of \cite{Nakayama_06}. Solution (II) includes a further 
D$_{13}$-state at $W$=2084 (and different parameter values for the other states
as summarized in Table II of \cite{Nakayama_06}). In further exploratory fits 
(Tables III - V in \cite{Nakayama_06}), more resonances were added, but did 
not improve the fit quality significantly. Therefore, in the present work we 
will only discuss solution (I).
A first analysis of the present quasi-free proton and neutron data was done 
in the following way. The results of the fits to the CLAS data for the free 
proton (solution (I)) of \cite{Nakayama_06}) have been adopted without any 
parameter change and folded with the momentum distribution of the bound proton, 
using the deuteron wave function in momentum space \cite{Machleidt_89}. 
They are then compared to the quasi-free proton results. For the neutron, 
only the electromagnetic photon - resonance couplings of all states have been 
varied, while all other resonance parameters (position, width, decay branching
ratios) have been taken from the proton fit. In a second fit, called solution
(Ia), an additional S$_{11}$ resonance was introduced because the neutron data
seems to show a broad bump at higher incident photon energies, which is not
apparent for the proton. In the following, we refer to solutions (I) and (Ia) 
as the NH model.

\section{Results}
\label{sec:results}

In the following all quasi-free differential cross sections are given in the cm 
(center-of-momentum) system of the incident photon and a target nucleon 
{\it at rest}. Apart 
from the immediate threshold region, such cross sections are only moderately 
smeared out by the effects of nuclear Fermi motion and can thus be compared 
almost directly to the corresponding results off free nucleons 
(see \cite{Krusche_95b} for details). The angular distributions have been 
fitted with Legendre polynomials 
\begin{equation}
\frac{d\sigma}{d\Omega} = \frac{q_{\eta '}^{\star}}{k_{\gamma}^{\star}}
\sum_{i} A_iP_i(cos(\Theta^{\star}_{\eta '})) \ ,
\end{equation}
where the $A_i$ are expansion coefficients. The phase-space factor 
${q_{\eta '}^{\star}}/{k_{\gamma}^{\star}}$ is also evaluated for the above 
cm system. The total cross sections have been extracted from the leading
Legendre coefficient $A_{0}$ of these fits. For some previous free 
proton results \cite{Williams_09,Crede_09}) only angular distributions but no 
total cross sections or total cross sections extracted by integration of the 
angular distributions have been given. For this reason, and in order to treat 
all data samples in a consistent way, also for these data sets total cross 
sections have been extracted in this work from the fits of the angular 
distributions. 

\subsection{Quasi-free proton cross section}

\begin{figure}[ht]
\resizebox{0.5\textwidth}{!}{%
  \includegraphics{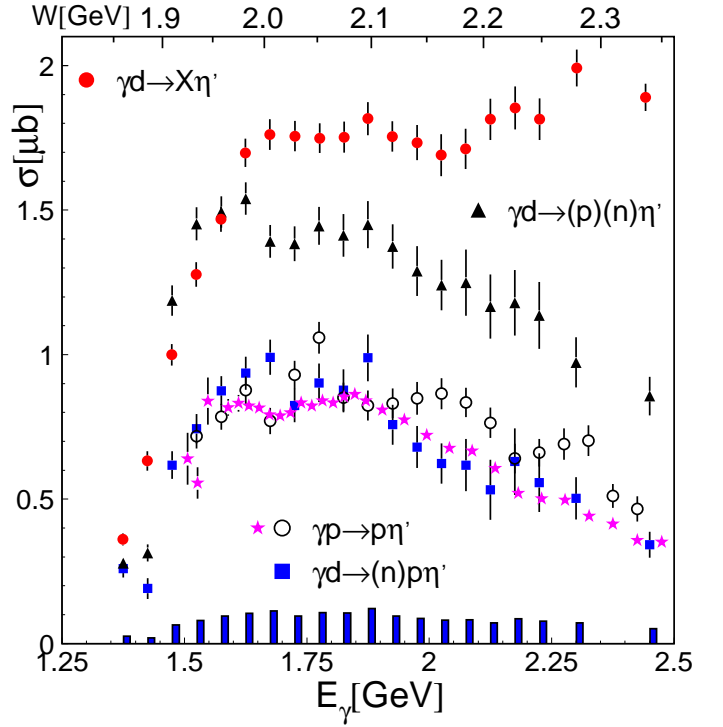}
}
\caption{Total cross sections for total inclusive $\gamma d\rightarrow X\eta '$
($\sigma_{x}$), inclusive quasi-free $\gamma d \rightarrow np\eta '$ ($\sigma_{np}$) and
quasi-free proton $\gamma d \rightarrow (n)p\eta '$ ($\sigma_p$) cross sections.
The quasi-free proton cross section is compared to the free proton results from
\cite{Crede_09} (open circles) and \cite{Williams_09} 
(magenta stars). 
Bar histogram at the bottom: systematic uncertainty of quasi-free proton data
excluding the overall normalization uncertainty.}
\label{fig:total_incl}       
\end{figure}

\begin{figure*}[th]
\resizebox{0.99\textwidth}{!}{%
  \includegraphics{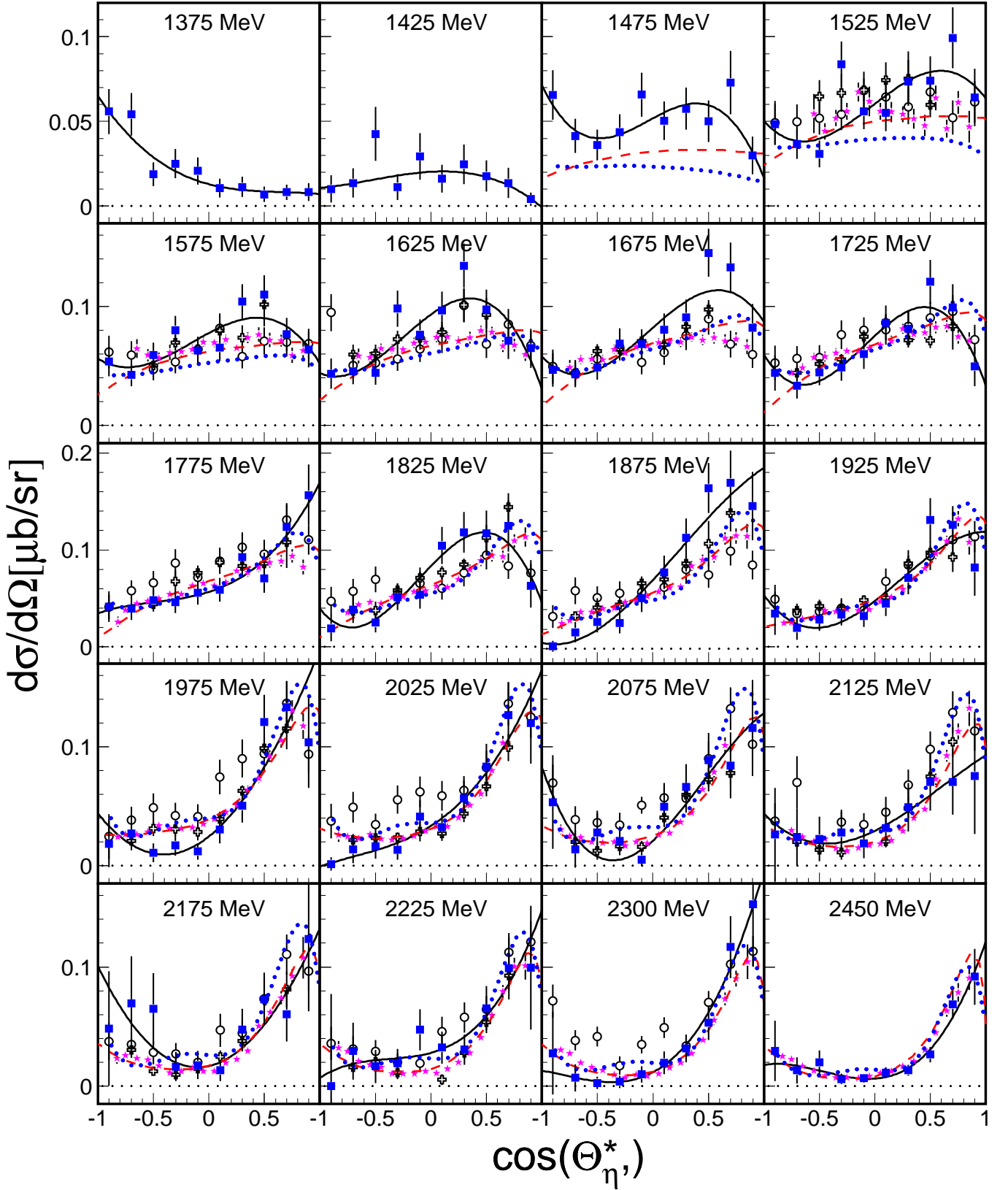}
}
\caption{Comparison of quasi-free $\eta '$ production off the bound proton
((blue) squares) to the free proton data: (black) open circles \cite{Crede_09},
(black) open crosses: \cite{Dugger_06}, (magenta) stars: \cite{Williams_09}. 
The numbers given in the figure indicate the bin centers in incident
photon energy (note: first two bins below free nucleon production threshold).
Note: results from \cite{Dugger_06,Williams_09} partly not exactly
for the same energy bins as present results. Closest bins or average of
overlapping bins chosen. All uncertainties only statistical.
Lines: Solid (black): Legendre fits to data present 
data, dashed (red): solution (I) NH model, dotted (blue): $\eta '$-MAID.
 }
\label{fig:proton}       
\end{figure*}

We first compare the results from the quasi-free $\gamma d\rightarrow (n)p\eta '$
reaction to free proton data. The angular distributions are summarized in Fig.
\ref{fig:proton}. The total quasi-free cross section off the proton is compared to
free proton results in Fig. \ref{fig:total_incl}. 
This figure shows also the inclusive quasi-free cross
section ($\sigma_{np}$) of single $\eta '$ production off the deuteron
(no condition on recoil nucleon) and the fully inclusive cross section 
($\sigma_x$) including contributions e.g. from $\eta '\pi$ final states. 
At the highest incident photon energies roughly 50 \% of the yield comes 
from such meson pairs. For the analysis of the single $\eta '$ channel
these multiple meson production reactions have been eliminated by the 
condition that no further mesons have been seen in the detector and by the 
kinematic constraints discussed in Sec. \ref{subsec:analysis} for events 
where additional mesons have escaped detection.

For the comparison of the quasi-free and free proton cross sections,
one could fold the free cross section data with the momentum distribution 
of the bound nucleons. However, simulations have shown, that for photon 
energies above 1.6 GeV this effect is small compared to the uncertainty of 
the data. Therefore we compare the unfolded data. Apart from a few energy 
bins close to threshold, the shapes of the angular distributions of quasi-free
and free proton data are in quite good agreement (see also Fig.~\ref{fig:coeff} 
for a comparison of the fitted Legendre coefficients). 
For the total cross section, shown in Fig.~\ref{fig:total_incl}, the agreement 
between the present quasi-free data (blue squares) and, in particular, the 
recent high precision proton data from CLAS \cite{Williams_09} 
(magenta stars) is excellent. The agreement with \cite{Crede_09} (open circles)
is within the systematic uncertainties. Altogether, no important nuclear 
effects were observed for quasi-free $\eta '$-production off the bound proton. 
Therefore, we expect that quasi-free $\eta '$-photoproduction off the bound 
neutron is a reasonable approximation of the free neutron reaction. 

\subsection{Inclusive quasi-free cross section}

The most simple approach to estimate the behavior of the neutron cross section 
is the measurement of the inclusive $\gamma d\rightarrow (np)\eta '$ reaction,
where only the $\eta '$ is detected, production of further mesons is excluded
by the missing mass cut, and no conditions for the detection of recoil nucleons 
are  applied. Since coherent contributions are small, the result is the incoherent 
sum of quasi-free proton and quasi-free neutron cross section.

\begin{figure*}[th]
\resizebox{0.99\textwidth}{!}{%
  \includegraphics{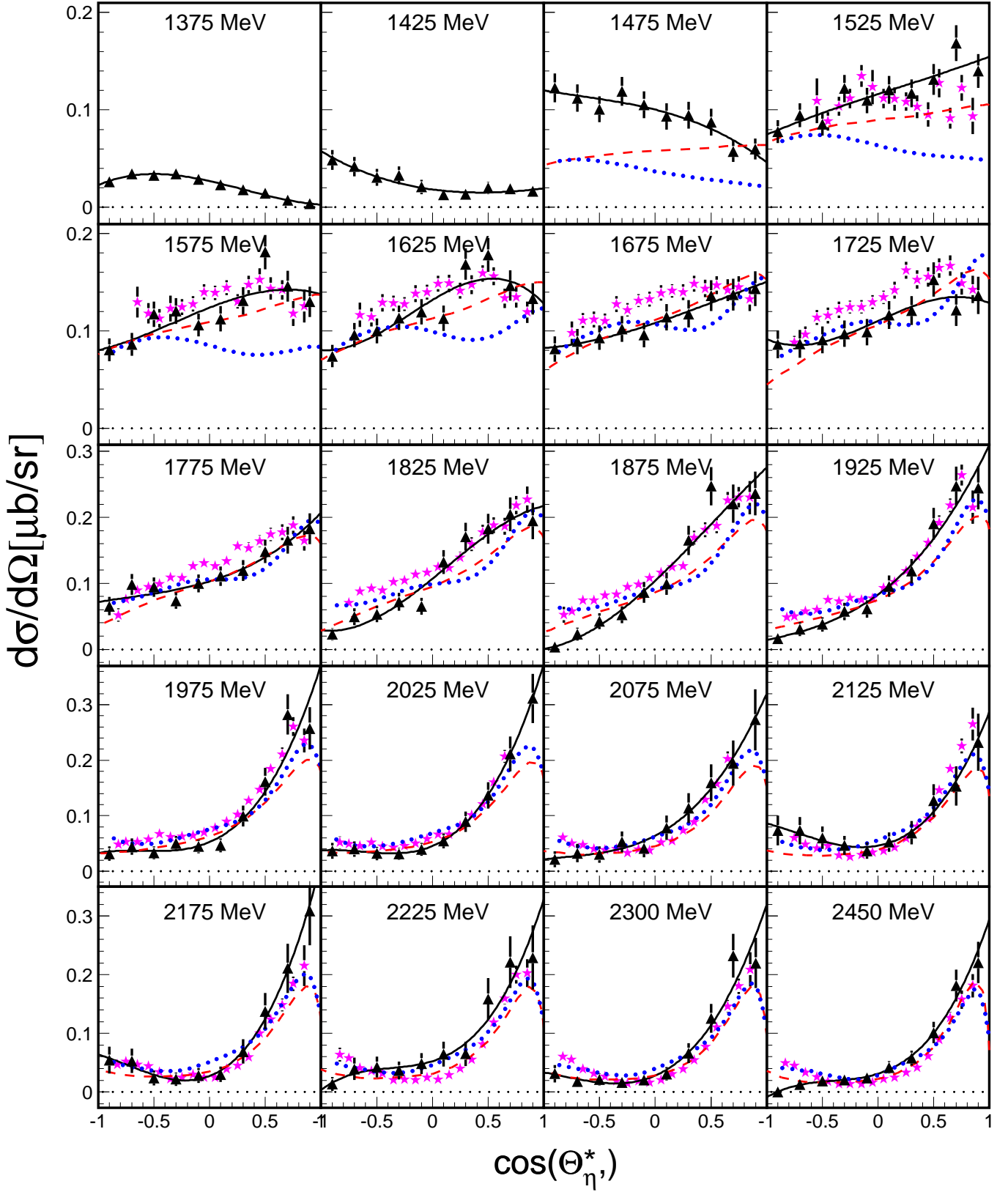}
}
\caption{Angular distributions for the inclusive quasi-free process $\gamma
d\rightarrow (np)\eta '$ of single $\eta '$ production (black triangles).
(Magenta) stars: free proton results from \cite{Williams_09} scaled up
by factor of two. 
Note: results from \cite{Williams_09} partly not exactly
for the same energy bins as present results. Closest bins or average of
overlapping bins chosen. All uncertainties only statistical.
Full (black)  lines: Legendre fit of present data, dashed (red) lines: solution (I) of 
NH model, (blue) dotted lines $\eta '$-MAID model. 
}
\label{fig:incl}       
\end{figure*}

The advantage of this approach is the comparably good statistical quality of 
data without detection of coincident neutrons. The angular distributions for 
this inclusive reaction are summarized in Fig. \ref{fig:incl} (the total cross 
section $\sigma_{np}$ is included in Fig. \ref{fig:total_incl}). 
The angular distributions are compared to the recent CLAS-data for the free 
proton \cite{Williams_09}, scaled up by a factor of two. 

\clearpage

\clearpage

\begin{figure}[ht]
\resizebox{0.5\textwidth}{!}{%
  \includegraphics{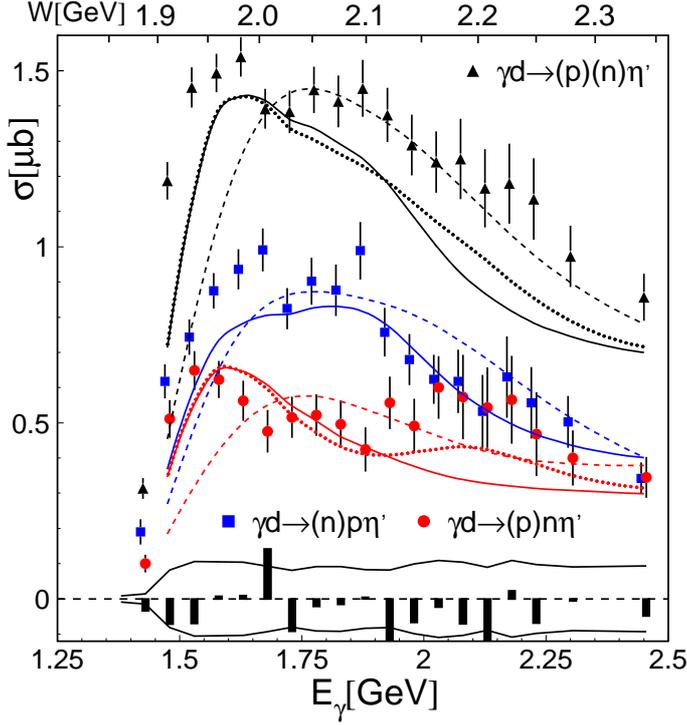}
}
\caption{Total cross section for inclusive  $\sigma_{np}$ , proton $\sigma_p$ , and 
neutron $\langle\sigma_n \rangle$ final state. 
Bar histograms: $(\sigma_n + \sigma_p - \sigma_{np})$, curves at bottom:
$\pm$ systematic uncertainty of $\sigma_n$. Proton (neutron) data points
slightly displaced to the left (right) for better readability of the figure. 
Curves: fits with reaction models model. NH model: solid: solution (I), 
dotted: solution (Ia), dashed: $\eta '$-MAID.  
}
\label{fig:total}       
\end{figure}

Agreement between the 
two data sets therefore signals regions where proton and neutron cross sections
are more or less identical. This is almost perfectly the case (sometimes with 
the exception of the extreme backward angles) for incident photon energies 
above 1.9 GeV. This is the region, where the angular distributions are 
strongly forward peaked, which in the models is mainly attributed to the 
contribution of $t$-channel processes.
At incident photon energies between 1.6 - 1.9 GeV, the inclusive cross section
is significantly smaller than twice the proton values, indicating a region, 
with $\sigma_n <\sigma_p$, which could be a first hint to different resonance 
contributions.
For the lowest energy bin (1475 MeV) effects of nuclear Fermi 
motion become important. Close to the threshold, energy conservation 
asymmetrically favors nucleon momenta anti-parallel to the incoming photon 
momentum, which results in an enhancement of meson backward angles 
(see \cite{Krusche_95b} for details). The data are compared to fits with the 
NH (solution (I)) and MAID model.
Shown is the incoherent sum of the model results for proton and neutron,
where the neutron couplings have been fitted to the quasi-free neutron data
(see next section). Agreement with the data for the other versions of the
NH model, which are not shown, is similar to solution (I).

\subsection{Quasi-free neutron cross section}

\begin{figure}[th]
\resizebox{0.5\textwidth}{!}{%
  \includegraphics{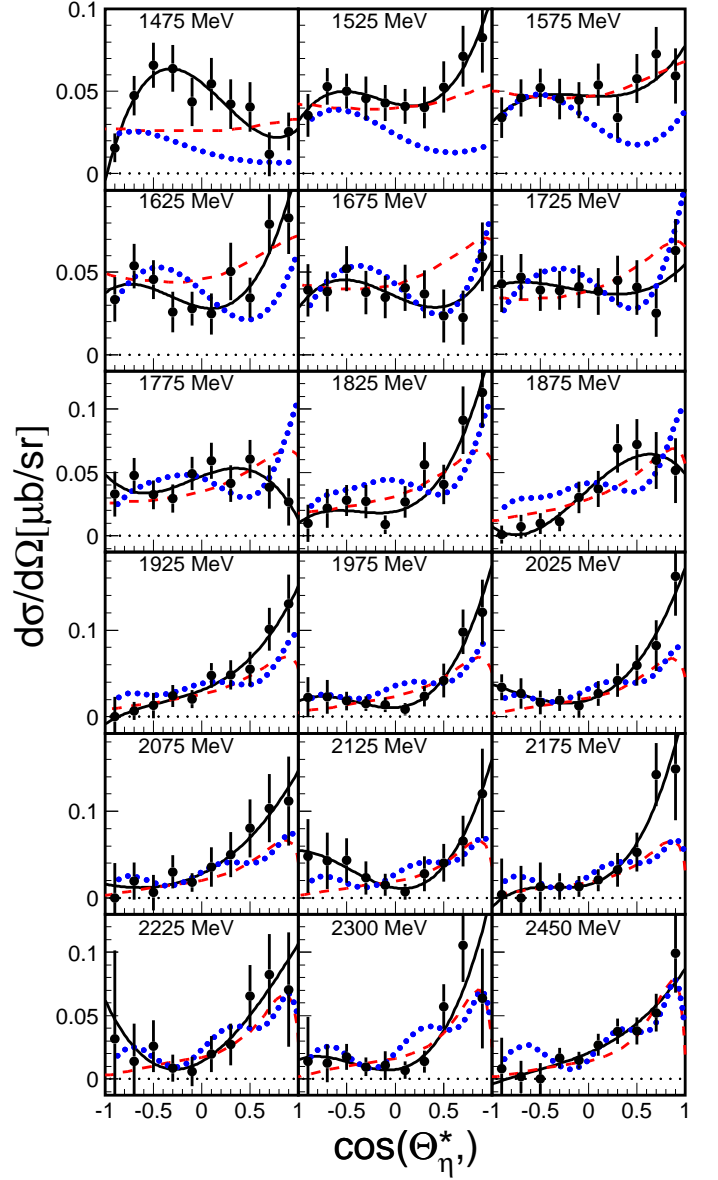}
}
\caption{Angular distributions for the quasi-free $\gamma n\rightarrow n\eta '$
reaction. Only statistical uncertainties.
Solid (black) lines: Legendre fit to data. Dashed (red) lines:
solution (I) of NH model, dotted (blue) lines: $\eta '$-MAID model.
}
\label{fig:neutron}       
\end{figure}

\begin{figure*}[tht]
\center{\resizebox{0.92\textwidth}{!}{%
  \includegraphics{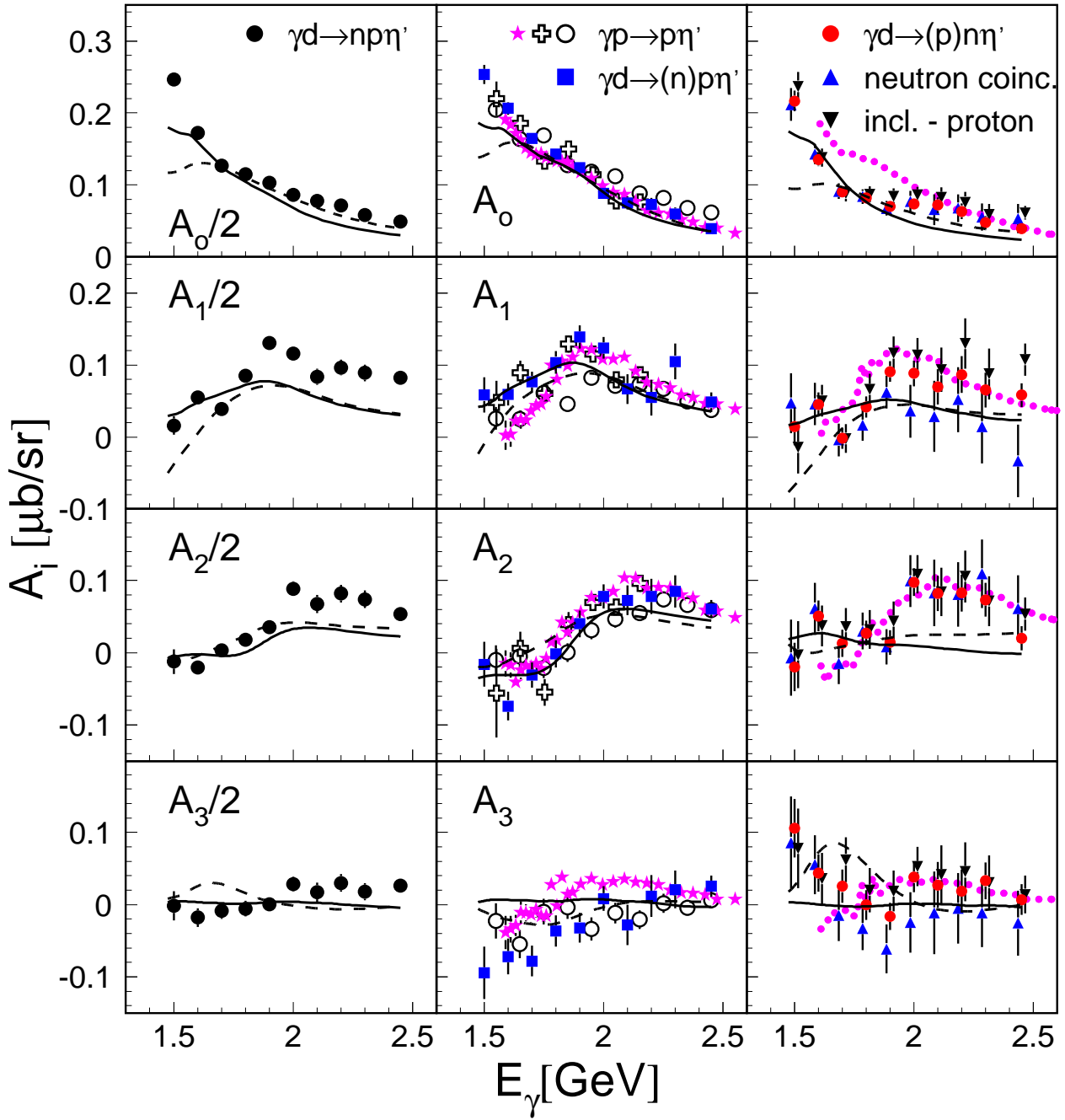}
}}
\caption{Coefficients of the Legendre Polynomials for the fitted angular
distributions. Left hand column: inclusive reaction scaled down by factor of 2.
Center column (proton targets): quasi-free data (blue squares);
free proton data: open crosses \cite{Dugger_06} (omitted for $A_{3}$)),
open circles \cite{Crede_09}, (magenta) stars \cite{Williams_09}. 
Right hand column (present quasi-free neutron data): (blue) upward triangles
from neutron coincidence, (black) downward triangles from difference of 
inclusive and proton data, (red) circles from averaged data. Symbols
slightly displaced to the left (right) for upward (downward) triangles
to make the plot better readable. 
In all plots solid lines: solution (I) NH model, dashed lines: $\eta '$-MAID;
for neutron: (magenta) dotted lines CLAS proton data.
}
\label{fig:coeff}       
\end{figure*}

As discussed in Sec. \ref{sec:analysis}, the quasi-free neutron cross section
can be extracted by coincident detection of the recoil neutrons or as 
difference of the inclusive and quasi-free proton cross section. The two 
methods give similar results, the averages of the angular distributions are 
summarized in Fig. \ref{fig:neutron}. For the fitted Legendre coefficients 
also the two individual data sets are shown in Fig. \ref{fig:coeff} as an 
estimate of systematic uncertainties.
The total cross section is compared in Fig. \ref{fig:total} to the proton data
and to the results from the reaction models. Differences between the two 
extraction methods are indicated by the bar histogram in this figure.

As expected from the discussion of the inclusive data, proton and neutron 
angular distributions are similar in magnitude and shape for photon energies 
above 1.9 GeV. However, at lower energies they are significantly different.
The high precision CLAS proton data \cite{Williams_09} show a kind of double
bump structure in the total cross section with shallow maxima around 
$W$= 1975 MeV and $W$=2080 MeV (cf. Fig. \ref{fig:total_incl}), which at the 
very limit of statistical significance is even reflected in the present proton 
and inclusive data  (cf. Fig. \ref{fig:total}). The earlier free proton 
CLAS data \cite{Dugger_06} may also contain such structures. In fact, 
the model calculation of Ref. \cite{Nakayama_06}, which fits the 
differential cross sections data of Ref. \cite{Dugger_06}, has predicted 
such shallow bump structures in the total cross section at about the same 
two energies. 

\begin{table*}
\renewcommand{\arraystretch}{1.25}
\caption{Resonance parameters of the $\eta^\prime\,MAID$ model \cite{Chiang_03}.
Resonance positions $M$ and total widths $\Gamma_{tot}$ in MeV.
Resonance couplings defined by $\chi^N_J \equiv \sqrt{\beta_{N\eta'}}A^N_J$
(in units of $10^{-3}GeV^{-1/2}$), where $\beta_{N\eta'}$ denotes the 
$N^*\to N\eta'$ branching ratio and, $A^N_J$, the helicity amplitude; 
$N=p,n$ and $J=1/2, 3/2$. $R$ is the neutron/proton ratio of the 
electromagnetic widths: $R\equiv\Gamma_{\gamma n}/\Gamma_{\gamma p}$. 
Coupling constants for the background nucleon born
terms $g_{\eta 'NN}=$-0.18 and vector meson exchange 
($\rho$, $\omega$), 
$g_{\rho NN}=$2.0, $\kappa_{\rho NN}=$3.5, 
$g_{\omega NN}=$12.0, $\kappa_{\omega NN}=$0.56 (vector coupling $g$ and 
tensor/vector ratio $\kappa$), $g_{\rho\eta '\gamma}=$1.24,
$g_{\omega\eta '\gamma}=$-0.43 (for details see \cite{Chiang_03}).}
\label{tab:maid}   
\begin{center}
\begin{tabular*}{\textwidth}
{@{\hspace{0.7cm}}c@{\hspace{0.7cm}}|@{\hspace{1.2cm}}c @{\hspace{1.2cm}}c
@{\hspace{1.5cm}}c @{\hspace{1.3cm}}c @{\hspace{1.3cm}}c
@{\hspace{1.3cm}}c @{\hspace{1.3cm}}c}
\hline\noalign{\smallskip}
Resonance & $M$ & $\Gamma_{tot}$ & $\chi^p_{1/2}$ &
$\chi^p_{3/2}$ & $\chi^n_{1/2}$ & $\chi^n_{3/2}$ &  $R$ \\
\noalign{\smallskip}\hline\noalign{\smallskip}
$S_{11}$  & 2004. & 286. &  19.7  &          & -14.6   &       &   -0.56   \\
$P_{11}$  & 2100. & 100. &  2.0   &          &   0.77  &       &   0.15  \\
$P_{13}$  & 1920  & 100  & -1.0   &  -4.2    &   5.0   & 0.97  &   1.4  \\
$D_{13}$  & 2150. & 230. &  12.0  &  -5.2    &  -1.4   & 0.96  &   0.02  \\
 \noalign{\smallskip}\hline
\end{tabular*}
\end{center}
\end{table*}

\begin{table*}
\renewcommand{\arraystretch}{1.25}
\caption{Resonance parameters of version (I) of the NH model \cite{Nakayama_06}. 
Notation as in Tab. \ref{tab:maid}.
Background parameters: $g_{\eta 'NN}=$0.43, 
$g_{\rho NN}=$3.3, $\kappa_{\rho NN}=$6.1, 
$g_{\omega NN}=$10.0, $\kappa_{\omega NN}=$0, $g_{\rho\eta '\gamma}=$1.25,
$g_{\omega\eta '\gamma}=$0.44. 
The spin-3/2 resonances, $P_{13}$ and $D_{13}$, are sub-threshold 
resonances and, as such, they may be considered as part of the background 
contribution. Further details of model 
(I) may be found in Table.I of Ref.~\cite{Nakayama_06}. Version (Ia) includes 
in addition a second S$_{11}$ resonance at $W$=2180 MeV and $\Gamma$=110 MeV.
} 
\label{tab:nh}   
\begin{center}
\begin{tabular*}{\textwidth}
{@{\hspace{0.7cm}}c@{\hspace{0.7cm}}|@{\hspace{1.2cm}}c @{\hspace{1.2cm}}c
@{\hspace{1.5cm}}c @{\hspace{1.3cm}}c @{\hspace{1.3cm}}c
@{\hspace{1.3cm}}c @{\hspace{1.3cm}}c}
\hline\noalign{\smallskip}
Resonance & $M$ & $\Gamma_{tot}$ & $\chi^p_{1/2}$ &
$\chi^p_{3/2}$ & $\chi^n_{1/2}$ & $\chi^n_{3/2}$ &  $R$ \\
\noalign{\smallskip}\hline\noalign{\smallskip}
$S_{11}$  & 1958. & 139. &  -12. &       & -17.  &       &   1.91   \\
$P_{11}$  & 2104. & 136. &  -13. &       &  -5.  &       &   0.16   \\
$P_{13}$  & 1885. &  59. &       &       &       &       &   0.02   \\
$D_{13}$  & 1823. & 450. &       &       &       &       &   1.24   \\
 \noalign{\smallskip}\hline
\end{tabular*}
\end{center}
\end{table*}

For the neutron, the first 
bump is more clearly visible, while the second one is suppressed.

The fitted coefficients of the Legendre representation of the angular 
distributions are summarized in Fig. \ref{fig:coeff}. For the quasi-free 
neutron the results for the two different extraction methods and for the 
average are shown.
Agreement of the two data sets for the even coefficients ($A_0$, $A_2$) is 
mostly  within statistical uncertainties. Some discrepancies outside 
statistical uncertainties are observed for $A_1$ and $A_3$. These odd 
coefficients are very sensitive to the extreme forward and backward angles, 
where due to small detection efficiencies uncertainties in the data are more 
important. For the same reason no values are given for $A_3$ for the earlier 
CLAS-data \cite{Dugger_06}.
Due to the relatively small angular coverage the fits were not sensitive to it.

The excellent agreement between free and quasi-free proton data is demonstrated 
in the center column of the figure. Effects from nuclear Fermi motion are 
mostly insignificant. The largest effect results again for the odd coefficients
($A_{1}$, A$_{3}$) since the asymmetric preference for nucleon momenta 
anti-parallel to the photon momentum induces a false forward-backward asymmetry 
in the quasi-free data. However, the effect is small. In case of the most
precise recent CLAS data, folding with Fermi motion (not shown in the figure)
improves slightly the agreement for the odd coefficients. 
For a better comparison of proton and neutron data, the new CLAS proton data
are included as magenta, dotted lines into the pictures of the neutron column.
The largest difference occurs for the $A_0$ coefficient, while the results 
are quite similar for the $A_1$, $A_2$, and $A_3$ coefficients. 
Only close to threshold there could be some systematic deviation.

The neutron data have been fitted with the NH model solution (I) (the
other solutions give very similar results) and the $\eta '$-MAID 
model. The results are shown in Figs.~\ref{fig:total}, \ref{fig:neutron}, 
\ref{fig:coeff}. 
For both models, all resonance parameters except the electromagnetic neutron
couplings were taken over from, respectively were dominated by, the fits to 
the proton data. For the NH model also a modified version with an additional 
S$_{11}$ resonance was tested, which, however did not much improve the 
agreement with the data. The parameters of the two models are summarized in
Tabs. \ref{tab:maid},\ref{tab:nh}. For both models agreement between fit and 
data is less good as for the proton data. For the total cross section, the 
MAID-fit does not well reproduce the threshold region. None of the fits 
reproduces the region above 2~GeV. For the coefficients of the angular 
distributions particular
$A_2$ disagrees with the data, which like for the proton rises above 2 GeV 
while it is flat and very small for the fits. 

In this most basic version both models include apart from the background terms
an S$_{11}$ and a P$_{11}$ resonance around $W$=2 GeV 
(respectively $W$=2.1 GeV). It must be emphasized again, that in both models 
these solutions are by far not unique. Nevertheless, there seems to be
agreement in so far, as both models need an S$_{11}$ resonance close to
threshold in order to fit the sharp rise of the total cross section and a
P$_{11}$ state to explain the shape of the angular distributions via an
S-P-interference term, which essentially has a linear cos($\Theta^{\star}_{\eta '}$)
behavior. Although also a P$_{13}$ state and/or the interference with 
background amplitudes can give rise to such a behavior. For both states there
are candidates in the Particle Data Group Review \cite{PDG}, the 
S$_{11}$(2090) and the P$_{11}$(2100) both one star states with not well 
defined
positions and widths. In case of the P$_{11}$ both model analyses result in
similar positions, widths, and neutron/proton ratio of the electromagnetic
couplings, although the absolute contribution of this state is stronger 
for the NH model. The S$_{11}$ has similar positions but different widths.
In case of the MAID model the proton coupling is stronger and there is a
negative sign between proton and neutron coupling, while for the NH model
the relative sign is positive and the neutron coupling is stronger.
Consequently, not even for these two `dominant' resonances agreement
is found between the two models. Obviously, further observables must be 
measured to arrive at better constraints for the model analyses. 

Here, one should also note, that already in the Lagrangian parameterizations 
of the background terms differences occur between the two models in both, the 
structure of some Lagrangians, as well as in the numerical values of coupling
constants. As an example for the vector meson currents, the coupling constant 
for the $t$-channel $\omega$-exchange is positive for the NH model, while it is 
negative for the MAID model (see Tabs. \ref{tab:nh},\ref{tab:maid}). On the 
other hand, the corresponding coupling for the $\rho$-meson is positive in both 
models. This leads to a destructive interference between this two terms in the 
MAID model, while the interference is constructive for the NH model. A more 
detailed analysis of the $t$-channel background terms is therefore also necessary. 
This might profit from data at higher incident photon energies, where this 
contribution dominates. Another example are the baryonic 
background currents where the NH model uses a pure pseudo-vector coupling
at the $NN\eta'$ vertex, while MAID uses pseudo-scalar coupling, again giving
rise to a relative sign between the amplitudes of the two models.  
  
\subsection{The coherent $\gamma d\rightarrow \eta ' d$ reaction}

\begin{figure}[th]
\resizebox{0.5\textwidth}{!}{%
  \includegraphics{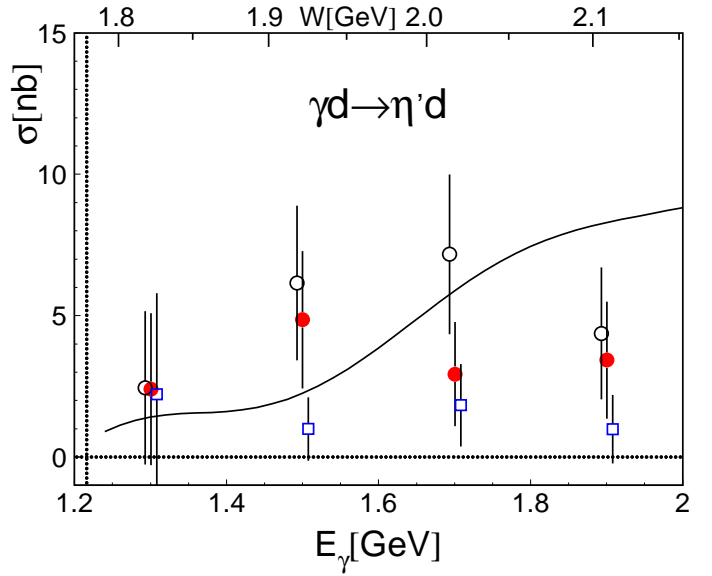}
}
\caption{Total cross section for the coherent reaction 
$\gamma d\rightarrow d\eta'$. Different symbols correspond to analyses with
different analysis cuts: closed (red) points: $\pm$25 MeV $\eta '$
missing mass, open circles: $\pm$50 MeV, open (blue) squares: -50 - 0 MeV.
Solid line: model prediction using the $\eta '$-MAID multipoles.
}
\label{fig:coh}       
\end{figure}

Coherent photoproduction is important due to its direct connection to the
iso-scalar parts of electromagnetic transition amplitudes. However, due to the 
dependence on the nuclear form factor it is strongly suppressed for
heavier mesons. The results for the total cross section obtained from the 
analysis discussed in Sec. \ref{sec:analysis} are summarized in 
Fig. \ref{fig:coh}. The typical size of systematic uncertainties can be
estimated by a comparison of the results from the three analyses using more
or less stringent cuts for the missing energy analysis, which is needed to
remove background from quasi-free processes. All results are on the order of 
only a few nb, the values from the two more stringent cuts ($\Delta E_{\eta '}$
between $\pm$25 MeV or between -50 - 0 MeV) are in reasonable agreement, those 
from the most open cut ($\pm 50$ MeV) may still include a small background
contribution. 

The results from the re-fitted reggeized MAID model \cite{Chiang_03} discussed 
above have been used to model the coherent reaction, using the parameters
summarized in Tab. \ref{tab:maid}. The formalism is based 
on a standard impulse approximation without multiple scattering contributions, 
i.e. the nuclear transition operator is taken as a sum of free single nucleon 
operators sandwiched between the deuteron wave functions. The general 
expression for the nuclear amplitude reads
\begin{equation}\label{10}
T_{M_fM_i\lambda_\gamma}=2\int\frac{d^3p}{(2\pi)^3}\
\psi_{M_f}^\dag\!\!\left(\vec{p}-\frac12\vec{q}\right)
t^{\lambda_\gamma}_{\gamma\eta^\prime}
\ \psi_{M_i}\!\!\left(\vec{p}-\frac12\vec{k}\right)\,,
\end{equation}
where $\vec{k}$ and $\vec{q}$ are momenta of the initial photon and a final 
meson. The indices $M_f$, $M_i$, and $\lambda_\gamma$ are respectively 
the $z$-projection of the final and initial deuteron spin and the photon 
polarization index. For the deuteron wave function, $\psi_M(\vec{p}\,)$, 
the momentum-space representation
\begin{eqnarray}\label{15}
\psi_M(\vec{p}\,) & = & \sum\limits_{L=0,2}
\sum_{M_L}(1M-M_L\, LM_L|1M)u_L(|\vec{p}\,|)\\
& & ~~~~~~~~~~~~~~~~~~~~~~~~~\times Y_{LM_L}(\hat{p})\chi_{M-M_L}
\nonumber
\end{eqnarray}
is taken, with $\chi_m$ being the triplet spin wave function.
For the spatial part $u_L(p)$ the parameterization \cite{Machleidt_87} derived
from the Bonn $NN$ potential model was used. For $\eta '$ photoproduction 
details of the deuteron wave function are more important than for lighter 
mesons since due to the large mass already at threshold large momentum 
transfers are involved. This causes a strong dependence on the behavior of the 
nuclear potential at small distances, in particular the $D$-wave component
of the wave function becomes important. The calculations predict, that for 
example at photon energies around 1.5 GeV the $D$-wave contributes already 
roughly 50 \% to the total cross section. 

The operator for $\eta^\prime$ photoproduction on a single nucleon has the 
well known spin structure in the Pauli matrix representation
\begin{equation}\label{20}
t_{\gamma\eta^\prime}^{\lambda_\gamma}=
K_{\lambda_\gamma}+\vec{L}_{\lambda_\gamma}\cdot\vec{\sigma}\,,
\end{equation}
with the spin-flip component $\vec{L}$ and the non spin-flip $K$.

The result corresponding to the parameter set from Tab. \ref{tab:maid} is 
compared to the data in Fig. \ref{fig:coh}. Given the simplicity of the 
impulse approximation, neglecting all two-body mechanisms as well as final 
state interaction effects, the agreement is quite good, demonstrating that 
the relative contribution of iso-scalar components is well represented in 
the model.
   
The predictions for the angular dependence are very sensitive to the assumed 
resonance and background contributions. For example, at low incident photon 
energies, where the S$_{11}$ resonance makes a large contribution, the 
spin-flip amplitude $\vec{L}$ is dominant giving rise to a forward peaking 
of the cross section. At higher incident photon energies, where other 
resonances and the $t$-channel background dominate, the spin-independent 
part $K$ is strong. This part is proportional to sin$(\Theta^{\star}_{\eta '})$ and 
thus vanishes at forward angles. Due to the statistical limitations of the 
data, it was not possible to extract angular distributions. A measurement with 
better statistical quality is highly desirable. 

\section{Summary and Conclusions}
\label{sec:conclu}
We have reported the first measurement of $\eta '$-mesons off the deuteron.
Both, the quasi-free and the coherent reaction carry important information
about the iso-spin composition of the elementary reaction on the free nucleon.

For the proton case, it has been demonstrated with this experiment, that the 
quasi-free cross section off the bound nucleon is very similar to the free
proton cross section. No significant nuclear effects e.g. from FSI processes
have been observed. At the given level of statistical precision of the data,
even effects from the momentum distribution of the bound nucleons are almost
insignificant. At low incident photon energies, they cause a small artificial 
forward-backward asymmetry in the angular distributions (coefficient $A_1$ 
in the Legendre series). At somewhat higher energies, folding the free proton 
cross section with the momentum distribution improves slightly (within 
systematic uncertainties) the agreement of free and quasi-free data for the
$A_3$ coefficient of the angular distributions. Only at very low incident 
photon energies, also the magnitude of the cross section is affected, but in 
that range free nucleon results are also not available or not precise. 
Agreement of the quasi-free data with the most recent measurement off the 
free proton from the CLAS-collaboration \cite{Williams_09} was found at a 
level much below the estimated systematic uncertainty of the present 
experiment. Consequently, the deuteron can be regarded as very well suited 
target to study the $\gamma n\rightarrow\eta ' n$ reaction.

The quasi-free reaction off the bound neutron has been studied in two different
ways with different sources of systematic uncertainty. In one approach, 
the $\eta '$-mesons were detected in coincidence with the participant neutrons.
In the second approach, the cross section obtained for coincident participant
protons was subtracted from the inclusive results without condition for recoil
nucleons. Since the detection efficiency for protons and neutrons is very 
different, a comparison of the two results gives a good estimate of the
systematic uncertainty for recoil particle detection. We had previously found 
excellent agreement for a similar analysis of $\eta$-photoproduction 
\cite{Jaegle_08}, which used the same data set. Also for the present analysis 
of $\eta '$-photoproduction good agreement is found, which indicates, that 
systematic effects are well under control. Only for the extreme forward and 
backward angular range, some discrepancies remain, which effect mainly the 
$A_{1}$ coefficient of the Legendre series for the angular distributions 
(see Fig. \ref{fig:coeff}). Altogether, the quality of the present 
$\gamma n\rightarrow n\eta '$ data is mostly limited by counting statistics, 
not so much by the systematic effects originating from the complications of 
a quasi-free reaction.

Proton and neutron cross sections behave similar at incident photon energies
above 2 GeV, where contributions from $t$-channel exchange are important.
At lower incident photon energies, in particular between 1.6 - 1.9~GeV, where
the proton cross section peaks, the behavior is different, which might point
to different resonance contributions, but could also arise from changing
interference terms between the resonances or between resonances and 
non-resonant background. 

The data have been compared to two different models, both with contributions
from a similar set of nucleon resonances and background terms, in
particular from $t$-channel mesonic currents. As already pointed out in
\cite{Nakayama_06} differential cross sections alone cannot uniquely determine
the contributing reaction mechanisms. Consequently, in the framework of
both models different solutions can be found. Future measurements of 
polarization observables have to clarify the situation. 

Finally, also a first estimate of the coherent $\gamma d\rightarrow d\eta '$
contribution at the level of at most a few nanobarn ($\sigma_{d\eta '}< 5$ nb
for all investigated photon energies is a reasonable estimate) has been 
extracted. This reaction is not only important for
the iso-spin separation of the elementary production amplitudes, but aims
also at the study of the $\eta '$ nucleon interaction via FSI effects.
The extracted results are consistent with an impulse approximation,
indicating, that the iso-spin composition of the model amplitudes is at least
not unreasonable and that FSI contributions are not substantial. However, the 
statistical limitation of the data is even more important at this low cross 
section level, so that effects beyond impulse approximation could not be 
studied.   
 
\vspace*{1cm}
\noindent{{\bf Acknowledgments}}

{\noindent We} wish to acknowledge the outstanding support of the accelerator group and operators
of ELSA. This work was supported by Schweizerischer Nationalfonds and Deutsche
Forschungsgemeinschaft (SFB/TR-16).


\begin{thebibliography}{99}
\bibitem{Krusche_03}     B. Krusche and S. Schadmand,     Prog. Part. Nucl. Phys.       {\bf  51},      399  (2003).
\bibitem{Burkert_04}     V.D. Burkert and T.-S. Lee,      Int. J. Mod. Phys. E          {\bf  13},     1035  (2004).
\bibitem{PDG}            C. Amsler et. al.,               Phys. Lett. B                 {\bf 667},        1  (2008).
\bibitem{CapRob}         S. Capstick and W. Roberts,      Phys. Rev. D                  {\bf  49},     4570  (1994);
                                                          ibid. D                       {\bf  57},     4301  (1998);  
							  ibid. D                       {\bf  58},   074011  (1998).
\bibitem{Krusche_95}     B. Krusche et al.,               Phys. Rev. Lett.              {\bf  74},     3736  (1995).
\bibitem{Krusche_97}     B. Krusche et al.,               Phys. Lett. B                 {\bf 397},      171  (1997).
\bibitem{Ajaka_98}       J. Ajaka et al.,                 Phys. Rev. Lett.              {\bf  81},     1797  (1998).
\bibitem{Bock_98}        A. Bock et al.,                  Phys. Rev. Lett.              {\bf  81},      534  (1998).
\bibitem{Armstrong_99}   C.S. Armstrong et al., 	  Phys. Rev. D  		{\bf  60},   052004  (1999).
\bibitem{Thompson_01}    R. Thompson et al.,		  Phys. Rev. Lett.		{\bf  86},     1702  (2001). 
\bibitem{Renard_02}      F. Renard et al.,		  Phys. Lett. B 		{\bf 528},	215  (2002).
\bibitem{Dugger_02}      M. Dugger et al.,		  Phys. Rev. Lett.		{\bf  89},   222002  (2002).
\bibitem{Crede_05}       V. Crede et al.,		  Phys. Rev. Lett.		{\bf  94},   012004  (2005). 
\bibitem{Nakabayashi_06} T. Nakabayashi et al., 	  Phys. Rev. C  		{\bf  74},   035202  (2006).
\bibitem{Bartholomy_07}  O. Bartholomy et al.,  	  Eur. Phys. J A		{\bf  33},	133  (2007).
\bibitem{Elsner_07}      D. Elsner et al.,		  Eur. Phys. J. A		{\bf  33},	147  (2007).
\bibitem{Denizli_07}     H. Denizli et al.,		  Phys. Rev. C  		{\bf  76},   015204  (2007).
\bibitem{Crede_09}       V. Crede et al.,		  Phys. Rev. C  		{\bf  80},   055202  (2009).
\bibitem{Williams_09}    M. Williams et al.,              Phys. Rev. C                  {\bf  80},   045213  (2009).
\bibitem{Sumihama_09}    M. Sumihama et al.,              Phys. Rev. C                  {\bf  80}, 052201(R) (2009).
\bibitem{McNicoll_10}    E.F. McNicoll et al.,            Phys. Rev. C                  {\bf  82},   035208  (2010). 
\bibitem{Ajaka_08}       J. Ajaka et al.,                 Phys. Rev. Lett.              {\bf 100},   052003  (2008).
\bibitem{Gutz_08}        E. Gutz et al.,                  Eur. Phys. J. A               {\bf  35},      291  (2008).
\bibitem{Horn_08a}       I. Horn et al.,                  Phys. Rev. Lett.              {\bf 101},   202002  (2008).
\bibitem{Horn_08b}       I. Horn et al.,                  Eur. Phys. J. A               {\bf  38},      173  (2008).
\bibitem{Kashevarov_09}  V.L. Kashevarov et al.,          Eur. Phys. J. A               {\bf  42},      141  (2009).
\bibitem{Gutz_10}        E. Gutz et al.,                  Phys. Lett B                  {\bf 687},       11  (2010).         
\bibitem{Nimai_95}       J.F. Zhang et al.,               Phys. Rev. C                  {\bf  52},     1134  (1995).
\bibitem{Ploetzke_98}    R. Pl\"otzke et al.,             Phys. Lett. B                 {\bf 444},      555  (1998). 
\bibitem{Chiang_03}      W.T. Chiang et al.,              Phys. Rev. C                  {\bf  68},   045202  (2003).
\bibitem{Dugger_06}      M. Dugger  et al.,               Phys. Rev. Lett.              {\bf  96},   169905  (2006). 
\bibitem{Nakayama_06}    K. Nakayama, H. Haberzettl,      Phys. Rev. C                  {\bf  73},   045211  (2006). 
\bibitem{Krusche_95b}    B. Krusche et al.,               Phys. Lett. B                 {\bf 358},       40  (1995).
\bibitem{Weiss_03}       J. Weiss et al.,                 Eur. Phys. J A                {\bf  16},      275  (2003).
\bibitem{Jaegle_08}      I. Jaegle et al.,                Phys. Rev. Lett.              {\bf 100},   252002  (2008).
\bibitem{Kuznetsov_07}   V. Kuznetsov et al.,             Phys. Lett. B                 {\bf 647},       23  (2007).
\bibitem{Miyahara_07}    F. Miyahara et al.,              Prog. Theor. Phys. Suppl.     {\bf 168},       90  (2007).
\bibitem{Husmann_88}     D. Husman, W.J. Schwille,        Phys. BL.                     {\bf  44},       40  (1988).
\bibitem{Hillert_06}     W. Hillert,                      Eur. Phys. J. A               {\bf  28},      139  (2006).
\bibitem{Elsner_09}      D. Elsner et al.,                Eur. Phys. J. A               {\bf  39},      373  (2009).
\bibitem{Aker_92}        E. Aker et al.,                  Nucl. Instr. and Meth. A      {\bf 321},       69  (1992).
\bibitem{Novotny_91}     R. Novotny,                      IEEE Trans. on Nucl. Science  {\bf  38},      379  (1991).
\bibitem{Gabler_94}      A.R. Gabler et al.,              Nucl. Instr. and Meth. A      {\bf 346},      168  (1994).
\bibitem{Suft_05}        G. Suft et al.,                  Nucl. Inst. Meth. A           {\bf 538},      416  (2005).
\bibitem{Mertens_08}     T. Mertens et al.,               Eur. Phys. J. A               {\bf  38},      195  (2008).
\bibitem{Bloch_07}       F. Bloch et al.,                 Eur. Phys. J. A               {\bf  32},      219  (2007).
\bibitem{Brun_86}        R. Brun et al.,                  GEANT, Cern/DD/ee/84-1,                 	     (1986). 
\bibitem{Zeitnitz_01}    C. Zeitnitz et al.,              The GEANT-CALOR
interface user's guide (2001)\\ (http://www.staff.uni-mainz.de/zeitnitz/Gcalor/gcalor.html).
\bibitem{Hejny_99}       V. Hejny et al.,                 Eur. Phys. J. A               {\bf   6},       83  (1999).
\bibitem{Schaefer_93}    E. Sch\"afer                     PhD thesis, University of Mainz, unpublished (1993). 
\bibitem{Nakayama_04}    K. Nakayama, H. Haberzettl.,     Phys. Rev. C                  {\bf  69},   065212  (2004).    
\bibitem{Lacombe_81}     M. Lacombe et al.,               Phys. Let. B                  {\bf 101},      139  (1981).    
\bibitem{Machleidt_89}   R. Machleidt                     Adv. Nucl. Phys.              {\bf  19};      189  (1989).
\bibitem{Machleidt_87}   R. Machleidt, K. Holinde, Ch. Elster, Phys. Rep.               {\bf 149},        1  (1987).   
\end{thebibliography}
\end{document}